\documentclass[a4paper,11pt]{article}
\pdfoutput=1 

\usepackage{jcappub} 

\usepackage[T1]{fontenc} 
\usepackage{subcaption}
\usepackage{tikz}
\usepackage{framed}

\def\d3k{{\displaystyle {\rm d}{\bf k} \over \displaystyle (2\pi)^3}}

\newcommand{\diff}       {\mm{\rm \,d}}

\newcommand{\Probability}{\mm{\rm Prob}}
\newcommand{\ProbDensity}{\mm{\cal P}}

\newcommand{\Matrix}     {\mm{\sf M}}
\newcommand{\ff}         {\mm{\bf f}}

\newcommand{\xx}         {\mm{\bf x}}
\newcommand{\yy}         {\mm{\bf y}}
\newcommand{\XX}         {\mm{\bf X}}
\newcommand{\rr}         {\mm{\bf r}}

\newcommand {\mm}[1] {\ifmmode{#1}\else{\mbox{\(#1\)}}\fi}

\title{\boldmath Stochastic Homology of \\
  Gaussian vs. non-Gaussian \\
  Random Fields:\\ \ \\
  Graphs towards Betti Numbers and Persistence Diagrams}
\author[a,b]{Job Feldbrugge}
\author[c]{Matti van Engelen,}
\author[b]{Rien van de Weygaert}
\author[d,b,e]{Pratyush~Pranav}
\author[c]{Gert Vegter}

\affiliation[a]{Perimeter Institute for Theoretical Physics,\\31 Caroline St. North, Waterloo, ON N2L 2Y5 Ontario, Canada}
\affiliation[b]{Kapteyn Astronomical Institute, University of Groningen,\\P.O. Box 800, 9747 AV Groningen, The Netherlands}
\affiliation[c]{Bernoulli Institute for Mathematics, Computer Science and Artificial Intelligence, University of Groningen, Nijenborgh 9, 9747 AG Groningen, the Netherlands}
\affiliation[d]{Univ Lyon, ENS de Lyon, Univ Lyon, CNRS, Centre de Recherche Astrophysique de Lyon UMR5574, FV69007, Lyon, France}
\affiliation[e]{Technion -- Israel Institute of Technology,\\Haifa, 32000, Israel}

\emailAdd{jfeldbrugge@perimeterinstitute.ca}

\abstract{The topology and geometry of random fields -- in terms of the Euler characteristic and the Minkowski functionals -- has received a lot of attention in the context of the Cosmic Microwave Background (CMB), as the detection of primordial non-Gaussianities would form a valuable clue on the physics of the early Universe. The virtue of both the Euler characteristic and the Minkowski functionals in general, lies in the fact that there exist closed form expressions for their expectation values for Gaussian random fields. However, the Euler characteristic and Minkowski functionals are summarizing characteristics of topology and geometry. Considerably more topological information is contained in the homology of the random field, as it completely describes the creation, merging and disappearance of topological features in superlevel set filtrations.

In the present study we extend the topological analysis of the superlevel set filtrations of two-dimensional Gaussian random fields by analysing the statistical properties of the Betti numbers -- counting the number of connected components and loops -- and the persistence diagrams -- describing the creation and mergers of homological features. Using the link between homology and the critical points of a function -- as illustrated by the Morse-Smale complex -- we derive a one-parameter fitting formula for the expectation value of the Betti numbers and forward this formalism to the persistent diagrams. We, moreover, numerically demonstrate the sensitivity of the Betti numbers and persistence diagrams to the presence of non-Gaussianities.}

\begin{document}
\maketitle
\flushbottom


\section{Introduction}
The Cosmic Microwave Background (CMB) radiation field is one of the principal cosmological probes, it is the source of revolutionary new insights into the
physics of the early Universe and the hitherto unimaginable accurate measurement of cosmological quantities that ushered the era of precision cosmology (see \citep{Cobe:1992, WMAP:2013, Planck:2018,Jones:2017}). The radiation field is a relic of an early dense phase of our Universe in which the Universe was filled with a hot photon-electron plasma. The primordial radiation field consists of a near perfect uniform radiation field with a blackbody spectrum with a temperature of $T_0= 2.725\text{K}$ \citep{Fixsen:1994, Fixsen:2009}. Perhaps the most outstanding property of this radiation field is the tiny temperature fluctuations with an amplitude in the order of $10^{-5}$. The statistical properties of these fluctuations represent a true
treasure trove of cosmological information, and form a direct reflection of processes that took place in the very early Universe. As evidenced by
a range of CMB observations, the CMB temperature
fluctuations and the corresponding mass density fluctuations, turn out to be close to a homogeneous and isotropic Gaussian random field \cite{smoot1992,bennett2003,
  spergel2007,Komatsu:2011,Planck:2015,Planck:2019}. The statistical properties of 
Gaussian random fields have been the subject of a range of fundamental studies \citep{Longuet-Higgins:1957,Adler:1981,Bardeen:1986,Adler:2009}. Statistically, 
Gaussian fields are completely described in terms of their autocorrelation function or power spectrum. Cosmological properties such as the curvature of space,
the energy content and the expansion rate determine the functional form of the power spectrum, so that one may infer constraints on these from
the CMB measurements.

Recent analyses of the Planck satellite data have set stringent limits on the level of possible deviations from Gaussianity of
the primordial density field \citep{Park:2004,Planck:2015,Planck:2019}. The detection of non-Gaussian perturbations will have major repercussions for our understanding of the
physical processes at the beginning of the Universe, in particular at its inflationary epoch \citep{guthpi1982,Baumann:2009,Chen:2010,Komatsu:2011,Bartolo:2012}.
There has been a considerable effort into determining the level and nature of these primordial non-Gaussianities, but these have not led
to any significant detection \citep{Planck:2019}. A considerable fraction of these studies is based on the measurement of the three-point and
four-point correlation function, or of their Fourier-space equivalents of the bispectrum and trispectrum. They concentrate on a restricted
set of deviations from Gaussianity, most notably the local, enfolded, equilateral and orthogonal templates for the bispectrum. 

Most of the present searches for primordial non-Gaussianities have been focused at detecting deviations of an a priori known form. In the light of the failure to
detect such signals, an arguably more profitable and objective approach would be to use a probe that is sensitive to generic non-Gaussian perturbations. We base
ourselves on the idea that the geometry and  topology of the primordial field is potentially one of the most direct and sensitive statistics, as it captures both the morphology, spatial distribution and connectivity between individual topological features in the primordial density field. It is these aspects that are very sensitive
to the nature of the random field \citep[see e.g.][]{Pranav:2019a,Pranav:2019c}. In fact, preliminary indications for this might be the mildly unusual
behaviour of the Euler characteristic pointed out by Eriksen et al. \citep{Eriksen:2004} and Park et al. \citep{Park:2004}.

Topology is the branch of mathematics that addresses the shapes, boundaries and connectivity of structural features
in a field. Traditionally the Minkowski functionals, which include the genus and the Euler characteristic, have been used to study the topology, morphology and geometry of
cosmological density fluctuations \citep{Doroshkevich:1970,Bardeen:1986,Gott:1986,Mecke:1994,Schmalzing:1997,Park:2013}. For the study
of Gaussian random fields, these topological and geometric measures have the virtue that there exist closed analytical expressions
for their expectation values as a function of the field's density level \citep{Adler:1981,Bardeen:1986,Tomita:1993,Schmalzing:1997, Appleby:2018, Pranav:2019a}.
While potential primordial non-Gaussianities might have expressed themselves via deviations from these statistical distribution
functions \citep{Pogosyan:2011,Codis:2013}, no significant signals have been detected \citep[see][]{Munshi:2012,Eriksen:2004, Park:2004, Modest:2013, Buchert:2017,
  Ducout:2013,Zhao:2014, Novaes:2016,Ganesan:2017}. Nonetheless, while the Minkowski functionals have been instructive, the topological information contained
in them is limited and convolved with geometric information. Moreover, they are not equipped to address the hierarchical aspects of the matter distribution directly.
Recent advances in the field of algebraic and computational topology \citep[see e.g.][]{Edelsbrunner:2002,Edelsbrunner:2009} have opened the possibility to
infer information on the contribution of individual features on the overall structural patterns, in terms of persistent topology and Betti numbers. 

\bigskip
In this study we therefore focus on the topological characterization of the cosmic mass (and temperature) distribution in terms of Betti numbers \citep{Betti:1871} and persistence diagrams \citep{Edelsbrunner:2002,Edelsbrunner:2009}. These are homology measures, concepts of algebraic and computational topology, describing in a quantitative manner how features in a manifold are connected through their boundaries \citep{Munkres:1984}. It concerns a language to assess the multiscale nature of the topology of the megaparsec cosmic mass distribution. {\it Betti numbers} are topological invariants that formalize the topological information content of the cosmic mass distribution by counting features in terms of the number of $p$-dimensional holes \citep{Edelsbrunner:1994,Zomorodian:2005,Robins:2006, Edelsbrunner:2009, Wasserman:2018}. For two-dimensional manifolds, the zeroth Betti number counts the number of disjoint components, while the first Betti number counts the number of inequivalent loops enclosing troughs. One often does not differentiate between the loops and the troughs they enclose. It is important to appreciate that the homological measures (with the exception of the Euler characteristic) are \textit{fundamentally non-local} as they depend on the correlation and connectivity of critical points in the random field. While homology and the Betti numbers do not fully quantify the topology of a manifold\footnote{One can change the topology without changing the homology of a manifold by removing a `hole' in one component of the manifold and introducing a `hole' in another connected component. The full topological information is captured in the homotopy. However, the homotopy is currently computationally out of reach.}, they extend the information beyond conventional cosmological studies of topology in terms of genus and Euler characteristics.

The profound significance of Betti numbers is underlined by their intimate relationship to the singularity structure and connectivity of the cosmic density field.
Van de Weygaert and collaborators \citep{Weygaert:2010,Weygaert:2011} introduced the concept of homology and Betti numbers in a cosmological context, in a study of
Betti number systematics in a range of weblike spatial galaxy
distributions. This was followed up by recent studies that invoked homology in a cosmological context along more systematic and formalized lines \citep{Weygaert:2011,
  Sousbie:2011a, Pranav:2017, Pranav:2019a}. The Betti numbers provide a summary of information on the topology of the cosmic mass distribution. A more detailed,
extensive and profound representation of related topological information is that of persistence \citep{Edelsbrunner:2002,Edelsbrunner:2009} 

Persistence describes the formation and demise of individual topological
features such as ``islands'', and ``troughs'' in the primordial density field, as we see them connecting and merging with surrounding topological features at different density
levels. Introduced by Edelsbrunner \citep{Edelsbrunner:2002}, persistence describes the density range over which topological features exist and hence reflects its
multiscale character\footnote{Note that persistence can be invoked for any scalar field.}. By focussing on the topological significance of features instead of their spatial scale, it provides a natural means of identifying structures
over a range of scales without having to invoke artificial filters. Adding to its significance is the fact that it is intimately coupled to the singularity structure of the
density field. This reflects the notion, stemming
from Morse theory \citep{Morse:1925, Milnor:1963}, that the existence of and connectivity between topological features is determined by the location and nature of critical
points -- {\it i.e.}, the maxima, minima and saddle points -- in the scalar field. In other words, persistence provides us with the mathematical language for
assessing and analyzing the hierarchy of topological structure in a density field.

Recent years have seen a proliferation of scientific studies invoking persistent topology to characterize the complexity of a large diversity of systems and
processes (see \citep{Wasserman:2018} for a recent review), ranging from brain research \citep{Petri:2014,Reimann:2017}, materials science \citep{Hiraoka:2016} to
  astrophysics and cosmology. Sousbie et al. \citep{Sousbie:2011a,Sousbie:2011b}, Shivashankar et al. \citep{Shivashankar:2016}, and Pranav et al. \citep{Pranav:2017}
  invoke persistence with the purpose to characterize the spine of the cosmic web
  \citep{Bond:1996,Weygaert:2008,Aragon:2010a,Cautun:2014,Libeskind:2018} and its connectivity structure. Recently, Kimura et al. \cite{Kimura:2016} determined persistence
  diagrams for (small) volume-limited samples of the DR12 release of the SDSS galaxy redshift survey in an attempt to characterize the topology of the
  spatial galaxy distribution, while Xu et al. \cite{Xu:2019} used persistence to identify individual voids and filaments in heuristic models of the
  cosmic matter distribution \citep[also see][]{Shivashankar:2016}. At a more fundamental level, Codis et al. \cite{Codis:2018} based their assessment of the connectivity of the nodes of the cosmic web on the persistent characterization of the cosmic web's spine. The concept of persistence and Betti numbers also offer a natural means of following the evolving topology of the reionization bubble network \citep{Elbers:2019}. In another astrophysical context, they were used to describe the topological structure of interstellar magnetic fields \citep{Makarenko:2018}. 
  
\bigskip
The main purpose of the present study is the formulation and inference of analytical descriptions and expressions of key topological characteristics, in particular
Betti numbers and topological persistence of two-dimensional random fields on $\mathbb{R}^2$. For an early report on this study we refer to \citep{bachelorThesis}. To this end,
we specifically concentrate on Gaussian random fields, and on possible small non-Gaussian deviations. The accompanying extensive numerical analysis of Gaussian field homology 
  is described in \citep{Pranav:2019a,Pranav:2019c}. While fully analytical expressions for the Euler characteristic and the Minkowski functionals
of Gaussian random fields exist \citep{Adler:1981,Bardeen:1986,Tomita:1993,Schmalzing:1997,Pranav:2019a}, to date no closed analytical expressions for topological quantities such as Betti numbers have been derived. All indications are that this may remain so \citep[see e.g.][]{Wintraecken:2013, Pranav:2019a}. While there are some high level, asymptotic results about Betti numbers for high density thresholds of Gaussian excursion sets in the mathematical literature, these are a consequence of the simple structure of Gaussian fields at these levels. As a consequence, nearly without exception, the information on Betti numbers is obtained indirectly, through numerical and statistical evaluations
\citep{Park:2013,Pranav:2019a}. The Gaussian Kinematic Formula (GKF) implies that the Minkowski functionals and the Euler characteristic can be expressed as integrals of local
functionals \citep[see][]{Adler:2009, Pranav:2019a}. This is not true for topological quantities, for which the localization is crucial. Note that the Euler characteristics
is an exception, as it can be written as the oscillating sum of the critical point densities. From a geometric perspective, while the Euler characteristic is topological, it can with the Gauss-Bonnet theorem be expressed in terms of a local characterization of the curvature. 

Motivated by the benefit of having an insightful and versatile analytical expression for Betti numbers and persistence diagrams, for Gaussian random fields, the present study follows an alternative route for inferring an accurate approximate formula. To this end, we follow a graph theoretical approach to Morse theory, developing path integral
expressions via the connectivity of singularities -- maxima, minima, and saddle points -- in a Gaussian random field \citep[also see][]{bachelorThesis}. While it is not
trivial to convert the Morse formalism into concise formulae such as entailed in the Gaussian Kinematic Formula \citep{Adler:2009}, in this study we demonstrate that
the numerically evaluated approximate expressions for the two-dimensional Betti numbers turns out to be remarkably accurate. 

\bigskip
Morse theory is the branch of mathematics that studies the singularity structure of a scalar field, \textit{i.e.}, the positions and connectivity of minima, maxima and saddle points. Of fundamental importance is Morse theory's principal topological tenet that there is a close relationship between the topology of the space and the critical points of any smooth function on the topological space \citep{Morse:1925, Milnor:1963,Edelsbrunner:2009}. Following this realization,
Morse theory describes the topology of the space by studying the critical points of a corresponding \emph{Morse function}, \textit{i.e}, a smooth scalar function
defined on the topological space with no degenerate critical points. Of particular interests are the observations that submanifolds defined as the regions where the Morse
function is in excess of a particular functional threshold value, \textit{i.e.}, superlevel sets, are topologically equivalent when the interval between the two
defining threshold values does not contain any critical point. The important implication of this is that all changes in the topology of a space occur only at
critical points. 

The close relation between the topology of a manifold and the spatial distribution of singularities finds its expression in the
\textit{Morse-Smale} complex \citep{Gyulassy:2008}. It is the segmentation of space defined by the spatial distribution of the singularities, consisting of regions
which connect minima and maxima via the field's integral lines. To formalize the connections between the singularities in the Morse-Smale complex, we introduce the concept
of the \textit{Morse Graph}. The graph is a diagrammatic representation of the connectivity structure of the Morse-Smale complex. We subsequently investigate the topological
structure of the field using the {\it incremental algorithm} \cite{Delfinado:1995}, which defines a filtration process to follow the addition and removal of topological
features as new singularities in the Morse graph are included. In combination
with the analytically known field distribution functions for the maxima, minima, and saddles in a Gaussian random field, we are led to the
integral expressions for the \textit{Betti numbers} as well as \textit{the persistence diagrams} of Gaussian random fields. By investigating
the asymptotic behaviour of the Betti numbers we are then able to infer an accurate fitting function. Note that these analytical expressions
have been derived on the basis of a probabilistic calculation, and do not involve and resort to realizations of Gaussian random fields. 

\bigskip
Having established accurate analytical expressions for the Betti numbers of Gaussian random fields, and the corresponding integral
expressions for persistence diagrams, we subsequently investigate the sensitivity of persistent Betti numbers and persistence diagrams on the presence of (subtle) non-Gaussianities in two-dimensional random fields. In this study, we restrict our numerical study to the local template version of the primordial non-Gaussianities and explore the changes
in Betti numbers and persistence diagrams as a function of the non-Gaussianity parameter $f_{NL}$ \citep[see][]{Komatsu:2011}. The results of this analysis
do underline the considerable potential for exploiting these instruments for a more profound topological data analysis of the cosmic microwave background -- and other
cosmological probes -- for assessing the presence of primordial non-Gaussianities. A few earlier studies already indicated this potential
\citep{Chingangbam:2012,ColeShiu:2018}. In fact, the recent study by Pranav et al.  \citep{Pranav:2019b} reports on the detection of possibly anomalous topological
signatures in the Planck CMB maps. 

\bigskip
In section~\ref{sec:randomfield}, we introduce basic concepts of (Gaussian) random fields, in particular focusing on the singularities in a random field and their statistical properties in a Gaussian random field. In section~\ref{sec:homology}, we discuss and describe the key concepts of homology, Betti numbers and persistence diagrams. In section~\ref{sec:graphform}, we embark on the statistical properties of Betti numbers, following the description of the Morse-Smale complex for Gaussian fields, the introduction of Morse graphs, and the use of the incremental algorithm in conjunction with Gaussian field singularity statistics to derive integral expressions for Betti numbers. We conclude this section with the derivation of an accurate analytical fitting formula for Betti numbers. Extending this analysis to the larger information content of the multiscale topological properties of Gaussian fields represented by persistence diagrams is the topic of section~\ref{sec:persistence}. Finally, we turn towards the potential of Betti numbers and persistence diagrams towards the detection of non-Gaussianities in section~\ref{sec:nongauss}. The conclusions of this study are summarized and discussed in section~\ref{sec:conclusion}.


\section{Random fields: definitions, probabilities \& singularities}
\label{sec:randomfield}
Throughout this paper we describe the topology of random fields, whereby we largely follow the notation introduced by
\cite{Bardeen:1986} \citep[also see][]{Weygaert:1996}. A typical example of a random field in the context of cosmology is
the three-dimensional density perturbation field 
\begin{equation}
\delta \rho(\textbf{x},t)\,=\,\frac{\rho(\textbf{x},t)-\rho_u(t)}{\rho_u(t)}\,,
\end{equation}
\noindent with $\rho({\bf x},t)$ the density at location $\xx \in \mathbb{R}^3$ at time $t$, and $\rho_u(t)$ the universal cosmological density for that cosmic epoch. In 
the present study we consider the primordial temperature fluctuations ${\delta T}(\textbf{x})$ of the current cosmic microwave background (CMB) temperature field $T(\textbf{x})$ at sky position
$\textbf{x} \in \mathbb{S}^2$ with respect to the current global CMB temperature $T_0=2.726 {\rm K}$ \citep{Fixsen:2009}, 
\begin{equation}
\delta T (\textbf{x})\,=\,\frac{T(\textbf{x})-T_0}{T_0}\,.
\end{equation}
The temperature field $T$ is a two-dimensional imprint of the density field $\rho$ at the time of last scattering when our Universe became neutral. As a consequence, the temperature perturbation $\delta T$ is a realization of a random field on the $2$-sphere $\mathbb{S}^2$. However, since our study of random fields is local, we for simplicity restrict the current analysis to random fields $[f]$ on Euclidean space $\mathbb{R}^2$. 

\medskip
Important for the cosmological context of the random fields is the \emph{statistical cosmological principle}, which
states that the statistical properties of the cosmic density distribution in our Universe are uniform throughout space
\footnote{Also crucial for the cosmological
reality is the {\it ergodic principle}. Based on this we are able to measure the value of ensemble averages by means of spatial
averages: these will be equal to the expectations over an ensemble of Universes. Given the fact that the Universe is unique, and its
density distribution the only realization we have of the underlying probability distribution, this is of key significance for the ability
to test theoretical predictions for stochastic processes like the cosmic mass distribution with observational reality. }. In
other words, the distribution function and moments of a field are the same in each direction and at each location, with the direct implication of the ensemble averages being dependent only on the distance between points.

To high accuracy, the primordial density perturbation $\delta \rho$, and the CMB temperature fluctuation field $\delta T$ --
which is a reflection of former -- are realizations of Gaussian random fields \citep{WMAP:2013, Planck:2018}, whose statistical character is fully specified by their first and second order moment. In this study, we largely focus on two-dimensional Gaussian random fields. However, the framework extends naturally to more general random fields which include non-Gaussianities. The detection or stringent bounds on primordial non-Gaussianities in the CMB are expected to lead to major improvements in our understanding of the early Universe \citep[see][]{2007JCAP...01..002C, 2014PDU.....5...75M, 2004JCAP...08..009B, 2005JCAP...06..003S, 2010AdAst2010E..72C, 2010JCAP...04..027C, 2005JCAP...09..011S, 2007JCAP...06..023C}.

\subsection{Gaussian random fields}
\label{sec:gaussian}
The random fields $f(\xx)$ studied in this paper are assumed to be smooth and continuous\footnote{In this section, the fields $f(\xx)$ may either be the raw unfiltered field or, without loss of generality, a filtered field $f_s(\xx)$.  A filtered field is a convolution with a filter kernel $W(\xx,\yy)$, $f_s(\xx)=\int \diff\yy f(\yy) W(\xx,\yy)$.}.
We generalize our study to any field $f(\xx)$ that is a linear functional ${\cal L}[g]$ of a random field $g(\textbf{x})$,
\begin{equation}
f(\xx)\,=\,{\cal L}[g;\xx]\,.
\end{equation}
Examples of such functionals are the value of the field itself at the point ${\xx}$, the gradient of the field $g({\xx})$, its Hessian, and more generally a convolution of
$g({\xx})$ with a kernel function $h({\xx})$,
\begin{align}
{\cal L}[g;\xx]&\,=\,g({\xx})\,;\qquad
{\cal L}_{i}[g;\xx]\,=\,\frac{\partial g}{\partial x_{i}} (\xx)\,;\qquad
{\cal L}_{i j}[g;\xx]\,=\,\frac{\partial^2 g}{\partial x_{i}\,\partial x_{j}} (\xx)\,;\label{eqn:functional}\\
{\cal L}_{\zeta}[g;\xx]&\,=\,\int\,h(\xx-\xx_\zeta)\,g(\xx_\zeta)\,{\rm d}{\xx_\zeta}\,.\nonumber
\end{align}
 
\noindent A random field $f$ is defined by its \emph{$N$-point joint probability},
\begin{align}
\ProbDensity_{\xx}( \ff) \diff \ff\,=\,\Probability\left[f(\xx_i)\in [f_i,f_i+\diff f_i],\ i=1,\ldots,N\right] \label{eqn:probability}\,,
\end{align}
the probability that a realization assumes a value in the interval $[f_i,f_i+\diff f_i]$, at the locations $\xx_i$ for $1 \leq i \leq N$. Throughout this paper, we use the shorthand $\xx=(\xx_1,\xx_2,\cdots,\xx_N)$ for the $N$ points $\textbf{x}_i$ and the vector $\ff = (f_1, f_2, \ldots, f_N)$ for the corresponding field values.

\emph{A Gaussian random field} is fully described by the first and second order moment, as can be seen from the probability distribution of a Gaussian random field with zero mean 

\begin{align}
  \ProbDensity_{\XX}  (\ff)  &\,=\,
{\displaystyle \exp \left[ {- \ff {\Matrix}^{-1} \ff^T /2} \right]  \over \displaystyle [(2 \pi)^N (\det \Matrix)]^{1/2}}\nonumber\\
&\,=\,
{\displaystyle \exp\left[-\frac{1}{2} \sum_{i,j}^N f_i (\Matrix^{-1})_{ij} f_j\right] \over \displaystyle [(2\pi)^N (\det \Matrix)]^{1/2}}\,,
  \label{eqn:distribution}
\end{align}

\noindent with $\Matrix^{-1}$ the inverse of the $N \times N$ covariance matrix $\Matrix$, \textit{i.e.},
\begin{align}
  M_{ij}  =  \langle f(\xx_i) f(\xx_j) \rangle\,= \xi(\textbf{x}_i - \textbf{x}_j)\,,
  \label{eqn:corrmat}
\end{align}
with the angle bracket denoting the ensemble average and the autocorrelation function $\xi(\textbf{r})$ \citep[see][]{Adler:1981, Bardeen:1986, Adler:2009}. The covariance matrix $\Matrix$ is the $N$-dimensional generalization of the variance $\sigma^2$ of the one-point Gaussian distribution,
\begin{align}
\sigma^2\,=\,\langle f(\xx) f(\xx) \rangle\,.
  \end{align}
The prefactor $ \left[(2 \pi)^N (\det \Matrix)\right]^{-1/2}$ in equation \eqref{eqn:distribution} normalizes the distribution, \textit{i.e.},
\begin{align}
\int   \ProbDensity_{\XX}  (\ff) \mathrm{d}^N\textbf{f} =1\,.
\end{align}
Given the fact that cosmological random fields obey the statistical cosmological principle, \textit{i.e.}, the fields are statistically isotropic and homogeneous, the correlation function
$\xi(\textbf{r})$ only depends on the distance $r = \| \textbf{r} \|$ between points,
\begin{equation}
\xi(r)\,=\,\xi(\textbf{r})\,.
\label{eq:xi}
\end{equation}
With respect to the higher order correlation functions, it is straightforward to infer from the distribution \eqref{eqn:distribution} that they can be expressed in terms of
the two-point correlation function. According to Wick's theorem, the odd correlation functions vanish, while the even correlation function are given by the sum of the
product of all possible pairings of points, \textit{i.e.}, 
\begin{align}
\langle f(\xx_1) f(\xx_2)f(\xx_3) \rangle &=0\,,\nonumber\\
\langle f(\xx_1) f(\xx_2)f(\xx_3)f(\xx_4)\rangle 
&=\xi(r_{12})\xi(r_{34}) + \xi(r_{13})\xi(r_{24}) + \xi(r_{14})\xi(r_{23})\,,\\
\langle f(\xx_1) f(\xx_2)f(\xx_3)f(\xx_4)f(\xx_5)\rangle &=0\,,\nonumber
\label{eq:wick}
\end{align}
with $r_{ij} = \|\textbf{x}_i - \textbf{x}_j\|$. 
Finally, it is important to note that when $f(\xx)$ is Gaussian random field, so too are the linear functionals ${\cal L}[f]$ of the field. \textit{I.e.} the gradient, Hessian and higher order gradients of a Gaussian field are themselves Gaussian fields.

\subsubsection{Multiscale Field Structure \& Fourier Components}
\noindent The {\it multiscale} structure of a Gaussian random field is most transparently described in terms of the Fourier components ${\hat f}({\bf k})$
of the field $f({\bf x})$, 
\begin{equation}
f({\bf x})=\int \d3k \ {\hat f}({\bf k})\,e^{-i{\bf k}\cdot{\bf x}}\,. 
\end{equation} 
In terms of the Fourier components of the field, the second order character of a Gaussian field is fully specified 
by the Fourier transform of the correlation function $\xi(r)$, known as the power spectrum $P(k)$,
\begin{align}
P(k) &= \int \xi(r)e^{-ikr}\frac{\mathrm{d}k}{2\pi}\,,
\end{align}
which can be interpreted as the two point function of the Fourier modes
\begin{align}
\langle {\hat f}({\bf k}) {\hat f}({\bf k}')\rangle\,&=\,(2\pi)^{3}\,P(k)\,\delta_D({\bf k}-{\bf k}')\,,
\end{align}
with the magnitude $k=\|\textbf{k}\|$ and the Dirac delta function $\delta_D$ \citep[see][]{Bertschinger:1992,Weygaert:1996}. Note that the correlation function in Fourier space is diagonal, \textit{i.e.}, $\langle \hat{f}(\textbf{k})\hat{f}(\textbf{k}')\rangle =0$ for $\textbf{k}\neq \textbf{k}'$.

The power spectrum $P(k)$ describes the prominence of the Fourier component at scale $\sim 2\pi/k$ in the field $f({\bf x})$, by means of
the square of the average amplitude of that component. Cosmologically speaking, this is a factor of central significance: 
the power spectrum fully specifies the nature of the primordial density and velocity field, out of which all structure in the cosmos emerges, and is itself a direct manifestation of processes in the very early (inflationary) Universe. 

\subsection{Non-Gaussian Random Fields}
A Gaussian random field is completely determined by the power spectrum or two-point correlation function. The higher order correlation functions can, with Wick's theorem, be expressed in term of the two-point function. For more general non-Gaussian random fields, the higher order correlation functions take a different form. The presence of these corrections are known as \textit{non-Gaussianities}. The most straightforward one is that
of a non-vanishing three-point function, and its Fourier transform, the bispectrum $B(k_1,k_2,k_3)$. The bispectrum is defined as
\begin{align}
\langle\hat{f}(\textbf{k}_1)\hat{f}(\textbf{k}_2)\hat{f}(\textbf{k}_3)\rangle &=(2\pi)^2 B(k_1,k_2,k_3)\delta^2(\textbf{k}_1+\textbf{k}_2+\textbf{k}_3)\,,
\end{align}
and provides a first measure of non-Gaussianities in the density perturbation field. Evidently, a non-Gaussian field may also have deviating higher order correlations. An immediate example is the four-point correlation function and its Fourier transform, the trispectrum $T(k_1,k_2,k_3,k_4)$, 
\begin{align}
\langle\hat{f}(\textbf{k}_1)\hat{f}(\textbf{k}_2)\hat{f}(\textbf{k}_3)\hat{f}(\textbf{k}_4)\rangle &=(2\pi)^2 T(k_1,k_2,k_3,k_4)\delta^2(\textbf{k}_1+\textbf{k}_2+\textbf{k}_3+\textbf{k}_4)\text{.}
\end{align}
To account for the Gaussian contribution to the four-point function, it is common to subtract the Gaussian four-point function from the trispectrum,
yielding the reduced trispectrum. In other words, a deviation of $T(k_1,k_2,k_3,k_4)$ from the value predicted by Wick's theorem indicates the presence of
non-Gaussianities.

\subsection{Critical Points}
\label{sec:critpnt}
The topological structure of a field $f$ is intimately connected to the character and distribution of the {\it singular} or {\it stationary points} 
in the field, as is illustrated by the Morse-Smale complex (see section \ref{sec:MorseSmale}). We identify the {\it critical} or {\it stationary points} in a differentiable field $f$ with the points for which the gradient of $f$ vanishes, \textit{i.e.}, 
\begin{align}
\nabla f \,=\,0\,.
\end{align}
Near a critical point $\textbf{x}_0$, the function $f(\textbf{x})$ can be approximate with the Taylor expansion
\begin{align}
\label{eq:taylor_exp_1}
f(\textbf{x})=&f(\textbf{x}_0) + (\textbf{x}-\textbf{x}_0)^T \nabla f(\textbf{x}_0) + 
 \frac{1}{2}(\textbf{x}-\textbf{x}_0)^T \mathcal{H}(\textbf{x}_0) (\textbf{x}-\textbf{x}_0) + \mathcal{O}\left(\|\textbf{x}-\textbf{x}_0\|^3\right)\\
             =&f(\textbf{x}_0) +
                \frac{1}{2}(\textbf{x}-\textbf{x}_0)^T \mathcal{H}(\textbf{x}_0) (\textbf{x}-\textbf{x}_0) + \mathcal{O}\left(\|\textbf{x}-\textbf{x}_0\|^3\right) \nonumber\,
\end{align} 
with the Hessian matrix field $\mathcal{H}$, whose components $\mathcal{H}_{ij}$ are the second order partial derivatives 
of the function $f$, \textit{i.e.},
\begin{align}
\mathcal{H}_{ij}= \frac{\partial^2 f}{\partial x_i \partial x_j}\,.
\end{align}
The function $f$ is said to be a Morse function if all its critical points are non-degenerate, \textit{i.e.}, the 
Hessian matrix $\mathcal{H}$ is non-singular in the critical points of $f$. According to Morse lemma, a Morse function $f$ is, in the neighbourhood of a critical point, approximately a quadratic form  
\begin{align}
f(\textbf{x})=f(\textbf{x}_0) +
                \frac{1}{2}\left(\lambda_1 y_1^2 + \lambda_1 y_1^2+ \ldots + \lambda_d y_d^2\right)+\mathcal{O}\left(\|\textbf{y}\|^3\right)\,,
\end{align}
with the coordinates $\textbf{y}=\textbf{x}-\textbf{x}_0=(y_1,y_2,\ldots,y_d)$ defined by the (normalized) eigenvectors of the Hessian matrix $\mathcal{H}$ and its non-zero eigenvalues $(\lambda_1,\lambda_2, \ldots, \lambda_d)$ \citep[see][]{Morse:1925, Milnor:1963}.

The character of a {\it singular point} is determined by the signature of the eigenvalues $(\lambda_1,\lambda_2, \ldots, \lambda_d)$, \textit{i.e.}, the number of 
positive and negative eigenvalues. The 
index $i$ of a singularity is defined as the number of negative eigenvalues. When all eigenvalues are negative, the singularity is a (local) {\it maximum} (index $i=d$). When all eigenvalues are positive, it is a (local) {\it minimum} (index $i=0$). The critical points for which the Hessian has a number of positive and negative eigenvalues are saddle points (index $0 < i < d$). In two dimensions, critical points with index $0$, $1$ and $2$ are minima, saddle points and maxima.

One of the implications of the Morse lemma is that non-degenerate singular points are isolated, \textit{i.e.}, two critical points of a Morse function are separated by a finite distance.

\subsubsection{Critical Point Densities}
The density of critical points of a random field, at function value $\nu=f/\sigma$, can be evaluated with Rice's formula \citep{Rice:1944, Longuet-Higgins:1957, Bardeen:1986, Adler:1981,Adler:2009}. For a two-dimensional random field, the density of minima $\mathcal{N}_0$, saddle points $\mathcal{N}_1$ and maxima $\mathcal{N}_2$ are given by the three-dimensional integral
\begin{align}
\mathcal{N}_i(\nu) =\iiint P(f=\nu \sigma, f_1=f_2=0,f_{11},f_{12},f_{13}) \begin{vmatrix} f_{11} & f_{12} \\ f_{12} & f_{22}\end{vmatrix}\mathrm{d}f_{11}\mathrm{d}f_{12}\mathrm{d}f_{22}\,,
\end{align}
with the notation
\begin{align}
f_i&=\frac{\partial f}{\partial x_i}\,,\quad f_{ij}=\frac{\partial^2f}{\partial x_i\partial x_j}\,,
\end{align}
for $i,j =1,2$. The integral for $\mathcal{N}_0,\mathcal{N}_1$, and $\mathcal{N}_2$ is performed over the configurations $(f_{11},f_{12},f_{22})$ corresponding to minima, saddle points and
maxima. Explicitly, the integral is performed over configurations $(f_{11},f_{12},f_{22})$ for which the eigenvalues $\lambda_1,\lambda_2$ of the Hessian $\mathcal{H}$, \textit{i.e.},

\begin{align}
\lambda_1 &= \frac{1}{2} \left[ f_{11}+f_{22} + \sqrt{4 f_{12}^2 +(f_{11}-f_{22})^2}\right]\,,\\
\lambda_2 &= \frac{1}{2} \left[ f_{11}+f_{22} - \sqrt{4 f_{12}^2 +(f_{11}-f_{22})^2}\right]\,,
\end{align}

\noindent are either both positive $\lambda_1 \geq \lambda_2>0$ for minima, of mixed signature $\lambda_1 > 0 > \lambda_2$ for saddle points, or  both negative $0>\lambda_1 \geq \lambda_2$ for the maxima. For statistically isotropic random fields, the three-dimensional integral can be reduced to a two-dimensional integral, by expressing the three second order derivatives $f_{11},f_{12},f_{22}$ in terms of two rotationally invariant variables. In terms of the two variables 

\begin{align}
J_1&=\lambda_1+\lambda_2=f_{11}+f_{22}\label{eqn:J1}\,,\\
J_2&=(\lambda_1-\lambda_2)^2=(f_{11}-f_{22})^2+4f_{12}^2\label{eqn:J2}\,,
\end{align}

\noindent as proposed by Pogosyan et al. \citep{Pogosyan:2011}, the number density $\mathcal{N}_i(\nu)$ of critical points in a Gaussian field of index $i$ takes the form

\begin{align}
  \mathcal{N}_i(\nu) = \iint \frac{|J_1^2-J_2|}{8\pi^2 \sqrt{1-\gamma^2}}\, \exp{\left[-\frac{1}{2} J_1^2 -J_2 - \frac{(\nu + \gamma J_1)^2 \sigma^2}{2(1-\gamma^2)}\right]}\,\,  \mathrm{d}J_1 \mathrm{d}J_2\,.
  \label{eqn:Ni}
\end{align}

\noindent In the expression above, we use the transformation of the measure
\begin{equation}
  \mathrm{d}f_{11}\mathrm{d}f_{12}\mathrm{d}f_{22} = \pi |\lambda_1-\lambda_2|\mathrm{d}\lambda_1\mathrm{d}\lambda_2\,.
\end{equation}
\noindent The integration domains for the minima, saddle points and maxima in the integral expression are given in
table \ref{table:integrationvariables}. The power spectrum, that characterizes the nature of the Gaussian random field,
influences the number densities via the spectral parameter $\gamma$ \citep[also see][]{Bardeen:1986}, 
\begin{align}
\gamma = \frac{\sigma_1^2}{\sigma_2 \sigma_0}
\end{align}
which is a combination of the spectral moments

\begin{align}
\sigma_i^2 = \int_0^\infty P(k) k^{2i + 1}\frac{\mathrm{d}k}{2\pi}\,.
\end{align}

\noindent The critical point densities of Gaussian random fields, and the subsequent analysis, is fully determined by the spectral parameter $\gamma$.

\begin{table}
\begin{center}
  \begin{tabular}{|p{3cm}||c||c|c|}
    \hline
    & & &  \\
    critical point & index $i$ & $J_1$ & $J_2$ \\
    & & &  \\
    \hline
    & & &  \\
   minima         &  $0$ & $(0,\infty)$ & $(0,J_1^2)$ \\ 
    & & &  \\
   saddle points     &  $1$ & $(-\infty,\infty)$ & $(J_1^2,\infty)$ \\ 
    & & &  \\
   maxima         &  $2$ & $(-\infty,0)$ & $(0,J_1^2)$ \\ 
    & & &  \\
   \hline
  \end{tabular}
  \caption{Field Hessian invariant variables $J_1$ and $J_2$: parameter domains for the number density of minima, saddle points, and maxima $\mathcal{N}_i$ in terms of the invariants $J_1$ and $J_2$ (for definitions see equations~\eqref{eqn:J1}, \eqref{eqn:J2} and \eqref{eqn:Ni}).}
\label{table:integrationvariables}
\end{center}
\end{table}

\begin{figure}
  \centering
  \mbox{\hskip 0.0truecm\includegraphics[width=0.85\textwidth]{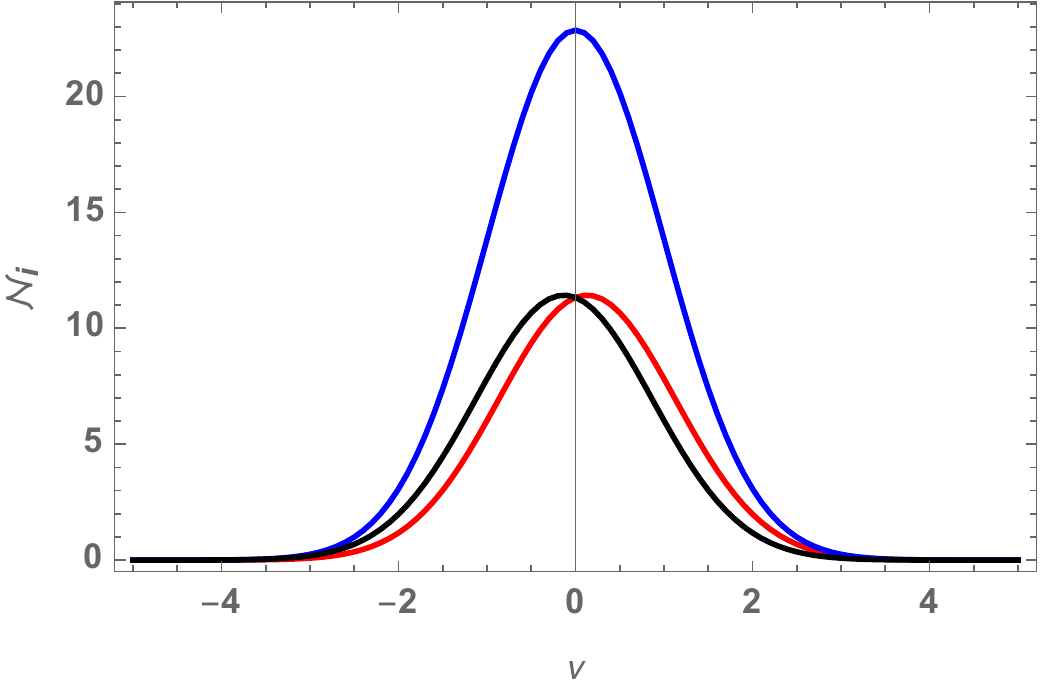}}
  \vspace{-0.25truecm}
  \caption{Critical point number densities $\mathcal{N}_i$ as a function of the field value (obtained using equation \eqref{eqn:Ni}). The number density of maxima (red), saddle points (blue) and minima (black) in a Gaussian random field (with LCDM power spectrum), at function value $\nu=f/\sigma$.}\label{fig:criticalPoint}
\end{figure}

\bigskip
\noindent In figure \ref{fig:criticalPoint}, we show the critical point densities $\mathcal{N}_0(\nu)$, $\mathcal{N}_1(\nu)$ and $\mathcal{N}_2(\nu)$ for
a two-dimensional Gaussian random field. Observe that for nearly the entire density range $\nu$ the saddle points are approximately twice as abundant as
the maxima and minima. The saddle point number density is symmetric and centers around the mean $\nu=0$. The number density of
maxima tends to be skewed towards positive field values, while the minima density is skewed towards negative field values. In Pranav et al. \cite{Pranav:2019a}, we analyze the number density of critical points in three-dimensional Gaussian random fields. In that situation, there is hardly any overlap between the number density
curves for minima and maxima. While in a two-dimensional random field both minima and maxima can be found for a large range of field values,
in three-dimensional Gaussian fields they hardly co-exist. This is related to the separation of homologies as demonstrated in \cite{Pranav:2019a}. Also note from figure~\ref{fig:criticalPoint} that for Gaussian random fields, the number densities of maxima and minima are symmetric around $\nu=0$, \textit{i.e.}, $\mathcal{N}_0(\nu)=\mathcal{N}_2(-\nu)$. For non-Gaussian random fields,
this symmetry is broken. 

\section{Homology and Persistence: \\ \ \ \ \ \ Topological Structure of Random fields}
\label{sec:homology}
The conventional analysis and characterization of random fields in terms of correlation functions inform us about their 
spatial coherence and clustering. In the case of Gaussian random fields we know that this is limited to the
two-point correlation function, which fully encapsulates the statistical nature of the field.

However, when the primary interest is that of the features and patterns that define the spatial structure field,
it is of greater benefit to study the topological characteristics of the field. Topology addresses the identity of
topological features and their spatial connectivity. These include islands, tunnels and cavities or voids. Islands are the regions that have a field value in excess of a specific threshold, and one may study their connection,
assess the number of cavities they encompass, how many tunnels percolate their interior, and a range of additional
questions of interest.

\bigskip
\noindent This section presents the key concepts for a topological analysis and description of
random fields that we concentrate on in the present study. These are homology, and specifically, the
aspects of Betti numbers and persistence. For the inference of analytical expressions for homology measures, which
is one of the key targets of the present study, we follow a path along a sequence of key topological concepts.
Starting with the definition of
a manifold, we are led to the concept of filtrations, in particular that of superlevel set filtrations. The
central step in our analysis is the intimate relationship between the topology of a manifold and its singularity
structure. The latter refers to the spatial distribution and connections between maxima, minima and saddle points.
It is one of the central tenets of {\it Morse theory}, and leads us to the concept of Morse-Smale complex for
codifying the connectivity between the manifold's singularities and its topological structure. 

The final, decisive and unique, element of our formalism concerns the graphical translation of the Morse-Smale complex,
producing a graphical representation of a random field realization that directly reflects its singularity
and topological structure. Mathematically, it is known as a {\it graph} \citep{Tutte:2001}. The graph representation
facilitates our ability to infer integral relations for the Betti numbers, and even for the persistence diagram.
While in principle we may follow this strategy in a three-dimensional context, in the present study we
restrict ourselves to the situation of a two-dimensional random field and show that it yields an approximate but
accurate analytical expression for the two Betti numbers $\beta_0$ and $\beta_1$. 

\subsection{Random Field Realizations and Manifolds}
The ultimate objective of our study, the Gaussian temperature fluctuation field of the CMB, concerns a 
realization of a random field on a $2$-sphere $\mathbb{S}^2$. However, because of the local nature of our analysis, we may restrict our analysis to random fields on the Euclidian plane $\mathbb{R}^2$. Hence, we consider a
function $f:\mathbb{R}^2\to \mathbb{R}$ that is a realization of a two-dimensional real random field
over $\mathbb{R}^2$. In principle, our analysis can be generalized to arbitrary higher dimensional
random fields. 

\bigskip
\noindent The realization $f(\rr)$ of the random field may be considered in terms of a manifold $M$. Roughly speaking, a manifold is
a subset of $\mathbb{R}^N$ which, for some $n\leq N$, locally looks like $\mathbb{R}^n$. For the field realization
$f(\rr)$, the corresponding manifold $M$ is defined as 

\begin{align}
M=\{(\textbf{r},f(\textbf{r}))| \textbf{r}\in \mathbb{R}^2\}\subset \mathbb{R}^3\,.
\end{align}

\bigskip
\subsection{Filtrations and Superlevel Sets}
On the manifold $M$ defined by the random field $f(\rr)$, we may construct a superlevel set. To this end, we apply a
threshold $\nu$ to the function values. The corresponding superlevel set filtration $M(\nu)$ of the manifold $M$ is 

\begin{align}
  M(\nu)=\{(\textbf{r},f(\textbf{r}))\in M|  f(\textbf{r})/\sigma\geq \nu\}\,.\label{eq:superlevelsetDef}
\end{align}

\begin{figure*}
\includegraphics[width=\textwidth]{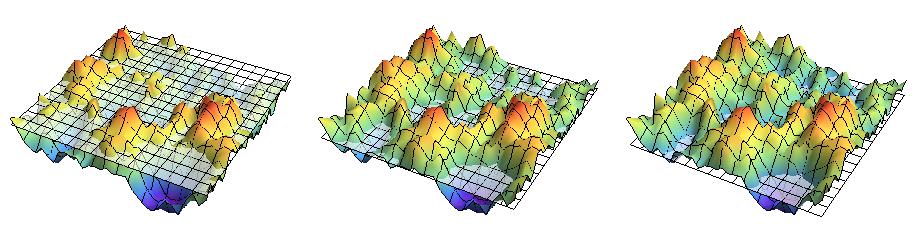}
\caption{Superlevel set filtrations. Superlevel set filtrations $M(\nu)$ of a realization of a Gaussian random field for three different field thresholds $\nu=f/\sigma$:
  the regions with field value above the threshold $f/\sigma>\nu$ belong to the superlevel set at $\nu$ (see equation \eqref{eq:superlevelsetDef}). }
\label{fig:GaussFieldmetNu}
\end{figure*}

\noindent A key property of superlevel set filtrations is that 

\begin{align}
M(\nu_1)\subseteq M(\nu_2) 
\end{align}

\noindent for all $\nu_1,\nu_2$ with $\nu_1 \geq \nu_2$. This can also be visually appreciated from the example of superlevel filtrations 
of a Gaussian random field realization at a sequence of three different threshold levels $\nu$ shown in figure~\ref{fig:GaussFieldmetNu}. 
In our analysis, $\nu$ always runs from $\infty$ to $-\infty$, incorporating more and more points of $M$ in $M(\nu)$ as we descend 
along $\nu$. By definition therefore, $M(\infty)=\emptyset$ and $M(-\infty)=M$.

\subsection{Homology of Random Fields:  Betti numbers}
\label{sec:betti}
Homology is the topological formalism for specifying in a quantitative and unambiguous way the connectivity of
manifolds through the connectivity of their boundaries \citep{Munkres:1984,Vick:1994,Edelsbrunner:2009}. A branch of
algebraic topology, it uses the mathematical formalism of homology groups to represent holes in a space in terms
of the connectivity with surrounding structures. We refer to appendix~\ref{appendix:Homology} for a formal treatment and
presentation of the basic concepts of homology \citep[also see][for complete mathematical treatments]{Rote:2006,Edelsbrunner:2009,Robins:2015}.
An extensive outline of homology, within the context of their application to the large-scale cosmological
mass distribution, can also be found in \cite{Weygaert:2011} and \cite{Pranav:2017}. Our recent numerical study of the Betti numbers,
Euler characteristic and Minkowski functionals of three-dimensional Gaussian fields \citep{Pranav:2019a}
can be seen as a companion paper to the present analytical and theoretical study.

\bigskip
\noindent Formally, Betti numbers are a concept from algebraic topology enabling a (partial) characterization of topological spaces.  
The Betti numbers characterize topology in terms of the homology of a manifold $M$, 
concentrating on the equivalence classes of homological chains and cycles on the manifold $M$ (or its simplicial decomposition). This is intimately related to a study of the boundary $\partial M$ of the manifold in the setting of superlevel set filtrations of functions on $\mathbb{R}^D$ studied in this paper.
The homology of a $D$-dimensional manifold is characterized
by $D+1$ Betti numbers $\beta_i, i=0,\ldots,D$.
In a sense, the $i^{\text{th}}$ Betti number may be considered as the number
of $i$-dimensional independent holes in a manifold $M$. For example, in three-dimensional space the zeroth Betti number $\beta_0$ counts
the number of separate components in $M$, while the first Betti number $\beta_1$ counts the number of tunnels and the second
Betti number $\beta_2$ the number of cavities, in the context of superlevel sets on $\mathbb{R}^3$.

A formal and rigorous definition of the Betti numbers is given in appendix~\ref{appendix:Homology}. It follows simplicial homology theory, and
concentrates on a characterization of a manifold in terms of the number of independent $p$-dimensional homological boundaries that it contains, each enclosing $p$-dimensional holes. Formally, within the context of algebraic topology, the boundary structure of
the manifold is assessed in terms of $p$-cycles (see app.~\ref{appendix:Homology}). The collection of independent $p$-dimensional homological cycles, up to topological equivalence, is the $p$-th {\it homology group} $H_{p}(M)$\footnote{Correctly defined, the $p$-th homology group is the $p$-th cycle group modulo the $p$-th boundary group.}. 
The rank of the homology group $H_{p}(M)$ is the number of all linearly independent topological cycles, and is denoted by the 
{\it Betti number $\beta_p$} \citep{Betti:1871}. 
A $D$-dimensional manifold $M$ has one homology group $H_p(M)$ for each of $D+1$ ranks $0\leq p \leq D$, and the set of homology groups characterizes the boundary structure and homology of the superlevel set.

We may also observe that as the number of $i$-dimensional holes is an integer this implies Betti numbers $\beta_i$
intrinsically to be an integer, \textit{i.e.}, ${\beta}_i:[-\infty,\infty]\to\mathbb{Z}$. For random fields over $\mathbb{R}^2$, a 
realization can have infinitely many $i$-dimensional holes. It is more sensible therefore to define normalized Betti numbers ${\tilde \beta_i}$,
the number of $i$-dimensional holes per unit volume/area. Hence, ${\tilde \beta_i}$ is a real number, \textit{i.e.},
${\tilde \beta}_i:[-\infty,\infty]\to\mathbb{R}$. For practical applications, such as the numerical evaluation
of Betti numbers in experimental or simulation datasets, de facto we nearly always deal with normalized Betti numbers. For
reasons of convenience, we therefore always retain the notation $\beta_i$, even when it concerns a normalized Betti
number $\tilde \beta_i$.

It is straightforward to extend the concept of the Betti numbers for a manifold to that of the superlevel set $M(\nu)$. 
The $i^{\text{th}}$ Betti number ${\beta}_i(\nu)$ of the superlevel set $M(\nu)$ is the number of $i$-dimensional holes
in $M(\nu)$ as a function of filtration level $\nu$. Dependent on the nature of the manifold, we may also identify symmetries with respect to its topological structure. The statistical symmetry of Gaussian random fields between underdense and overdense regions implies an equal number of overdense regions above level $\nu$ versus the number
of underdense troughs below level $-\nu$. This means that in two-dimensional space the Betti numbers obey
the statistical symmetry relation

\begin{align}
\beta_0(\nu)=\beta_1(-\nu)\,. 
\end{align}

\bigskip
  
\subsection{Betti, Euler, and Genus}
\label{sec:euler}
\bigskip
The homology and connectivity of a manifold are globally quantified by the Betti numbers, which represent a (partial) characterization of
topological spaces\footnote{Strictly speaking this is only true when ignoring 
  the torsion-free part of a homology group}. In this sense, they extend the principal topological characterization known in cosmology in terms
of {\it genus} or {\it Euler characteristics}. Numerous cosmological studies have assessed and discussed the {\it genus} of the isodensity
surfaces defined by the Megaparsec galaxy distribution \cite{Gott:1986,Hamilton:1986,Hoyle:2002}.

The genus $G$ specifies the number of handles defining an orientable surface, \textit{i.e.}, the maximum number of incisions one can make without partitioning the manifold into separate components. The genus has a direct
and simple relation to the Euler characteristic $\chi$ of an isodensity surface.

Originally stemming from the description of polyhedral surfaces and three-dimensional simplicial complexes, the Euler characteristic
is a topological invariant that plays a key role in homology.  For a polyhedral surface with $V$ vertices, $E$ edges and $F$ faces, the
Euler characteristic is
\begin{equation}
\chi\,=\,V-E+F\,.\label{eq:verg}
\end{equation}
This definition of the Euler characteristic extends to manifold, as any manifold has a polyhedral decomposition. Note that this definition is well-defined, as the Euler characteristic does not dependent on the chosen decomposition.
While in principle one may define the Euler characteristic in any dimension $D$, its significance may be best appreciated in the context of surfaces in 3-dimensional space. To this end, we consider a 2-manifold in three-dimensional space, the surface
${\cal S}={\partial M}$ of a 3-manifold $M$. It follows from equation \eqref{eq:verg} that, when the surface ${\cal S}$ consisting of $c$ discrete components, the genus of the surface is given by
\begin{equation}
G\,=\,c\,-\,{1 \over 2}\,\chi({\cal S})\,,
\end{equation}
where the Euler characteristic $\chi({\cal S})$ is, by the Gauss-Bonnet theorem, expressed as the integrated intrinsic curvature of the surface 
\begin{equation}
\chi({\cal S})\,=\,{\displaystyle 1 \over \displaystyle 2\pi}\,\oint\,\left({\displaystyle 1 \over \displaystyle R_1 R_2}\right)\,dS\,. 
\label{eq:euler}
\end{equation}
Here $R_1$ and $R_2$ are the principal radii of curvature at each point of the surface, and the integrand $1/R_1R_2$ is known as the Gaussian curvature.
The integral of the curvature is invariant under continuous deformation of the surface ${\cal S}$. This is perhaps one of the most surprising results in
mathematics, establishing an an intricate and profound relation between differential geometry and topology.

\bigskip
\noindent The profound central significance of the Euler characteristic is not only reflected in its intrinsic and deep
relationship with the topology of simplicial complexes and the geometric properties of manifolds, and hence with simplicial
topology and differential geometry. Similar intricate relations exist between the Euler characteristic and the homology
of a manifold, and also with the singularity structure of manifolds. This is reflected in the relation between the
Euler characteristic $\chi(M)$ of a manifold and its Betti numbers $\beta_i(M)$.

\begin{figure}
  \centering
  \mbox{\hskip 0.0truecm\includegraphics[width=0.8\textwidth]{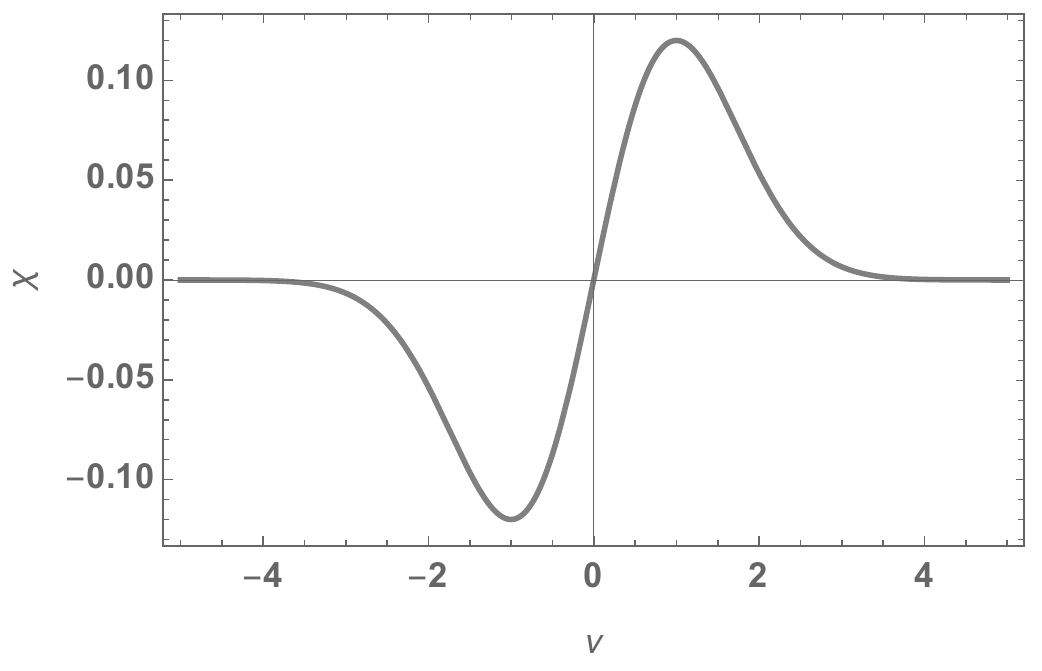}}
  \vspace{-0.25truecm}
  \caption{The Euler characteristic $\chi(\nu)$ for the superlevel set filtration of
    a Gaussian random field on a $\mathbb{R}^2$ at field threshold $\nu=f/\sigma$. The well-known analytical expression for $\xi(\nu)$ is given in
    equation~\eqref{eqn:chigauss}.}
   \label{fig:chi}
\end{figure}

The {\it Euler-Poincar\'e formula} \cite{Edelsbrunner:2009} establishes that for any arbitrary $D$-dimensional manifold $M$, its Euler characteristic \footnote{For the $D$-dimensional definition of the Euler characteristic see \cite{Edelsbrunner:2009}.}
$\chi(M)$ is the alternating sum of the corresponding Betti numbers $\beta_p(M)$ of the manifold, 
\begin{equation}
\chi\,=\,\sum_{p=0}^D\,(-1)^p \beta_p\,.
\end{equation}
In addition, according to Morse theory (see sect.~\ref{sec:graphform}) there is an intimate link between the topology of a field and its singularity
structure. It expresses itself, amongst others, in the relation between the number density of critical points and the Euler characteristic $\chi(\nu)$.
Akin to the relations above, the Euler characteristic $\chi(\nu)$ of the manifold filtration at threshold level $\nu$ appears to be equal to the alternating sum of all critical points with a field value in excess of $\nu$,

\begin{align}
  \chi(\nu) &= \int_{\nu}^\infty (\mathcal{N}_0(\nu')- \mathcal{N}_1(\nu') + \mathcal{N}_2(\nu'))\mathrm{d}\nu'\,.
\end{align}

\noindent As calculated by \cite{Bardeen:1986}, for a Gaussian field this implies the following analytical expression for
the Euler characteristic at level $\nu$,

\begin{align}
\chi(\nu)\,=\, \left(\frac{(2\pi)^{3/2}}{2} \,\frac{\sigma_1^2}{\sigma_0^2}\right)\,\,\, \nu\ e^{-\nu^2/2}\,.
\label{eqn:chigauss}
\end{align}

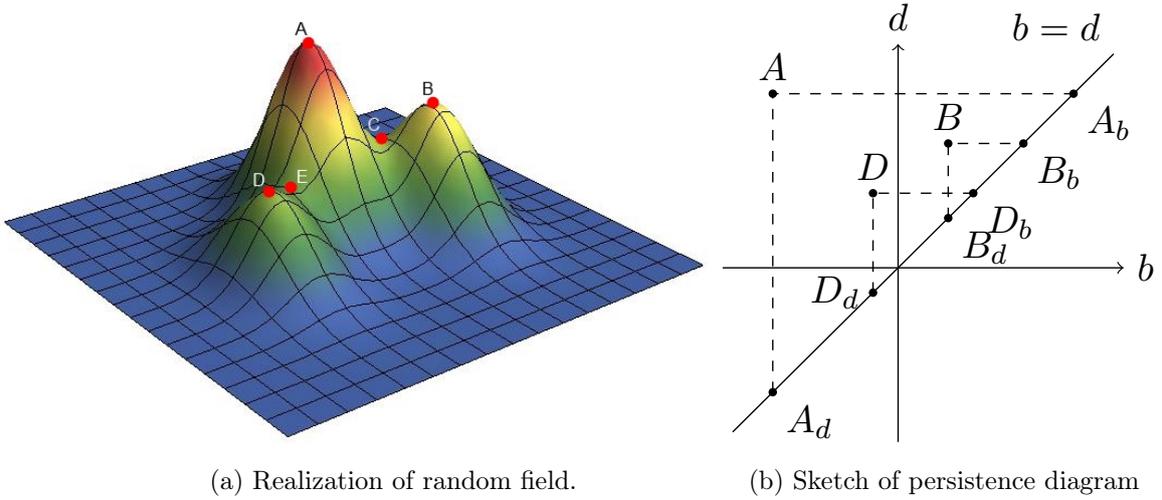
\begin{figure}
\centering
\vskip -0.75truecm
\begin{subfigure}[b]{0.65\textwidth}
\mbox{\hskip -0.5truecm\includegraphics[width=\textwidth]{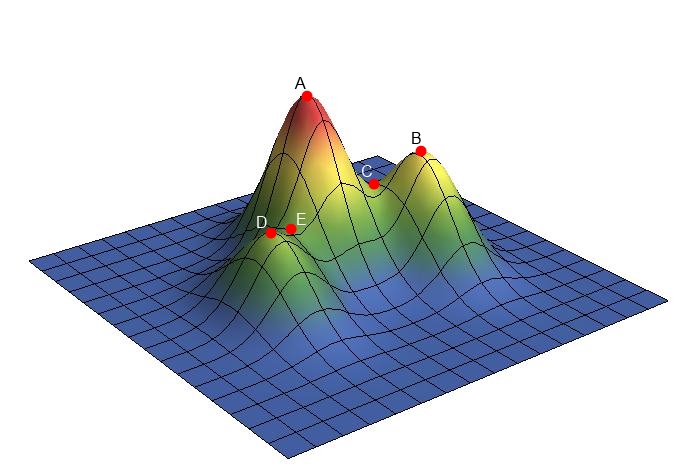}}
\caption{Realization of random field.}
\label{fig:PersExp1}
\end{subfigure}
\mbox{\hskip -1.0truecm
\begin{subfigure}[b]{0.4\textwidth}
\resizebox {\textwidth}{!}{
\begin{tikzpicture}[scale=1.0]
\draw[->] (-2,-0.25) -- (2,-0.25) node[right] {$b$};
\draw[->] (-0.25,-2) -- (-0.25,2) node[above] {$d$};
\draw (-1.9,-1.9) -- (1.9,1.9) node[above left] {$b=d$};
\draw[dashed] (-1.5,-1.5) -- (-1.5,1.5) -- (1.5,1.5);
\draw[dashed] (-0.5,-0.5) -- (-0.5,0.5) -- (0.5,0.5);
\draw[dashed] (0.25,0.25) -- (0.25,1) -- (1,1);
\filldraw  (-1.5,1.5) circle (1pt) node[above] {$A$};
\filldraw  (-0.5,0.5) circle (1pt) node[above] {$D$};
\filldraw  (0.25,1) circle (1pt) node[above] {$B$};
\filldraw  (-1.5,-1.5) circle (1pt) node[below right] {$A_d$};
\filldraw  (1.5,1.5) circle (1pt) node[below right] {$A_b$};
\filldraw  (-0.5,-0.5) circle (1pt) node[left] {$D_d$};
\filldraw  (0.5,0.5) circle (1pt) node[below right] {$D_b$};
\filldraw  (0.25,0.25) circle (1pt) node[below right] {$B_d$};
\filldraw  (1,1) circle (1pt) node[below right] {$B_b$};
\end{tikzpicture}
}
\caption{Sketch of persistence diagram}
\end{subfigure}}
\caption{Persistence and Field singurlarity structure: Panel (a) illustrates a realization of a random field. The critical points are denoted by $A,B,C,D$ and $E$. The corresponding superlevel set filtrations are plotted in the six lower panels. Panel (b) illustrates the persistence of the superlevel set filtrations of the realization shown in the left figure. The persistence diagram consists of the points $A$, $B$ and $D$, corresponding to the birth-death pair or the corresponding critical point in the realization. $A_b$, $B_b$ and $D_b$ are the times of birth corresponding to these critical points, while $A_d$, $B_d$ and $D_d$ are the corresponding times of death. 
  Panels bottom: Persistence and Field singurlarity structure. A top view of the superlevel set filtration $M(\nu)$ for decreasing
  threshold $\nu$.}
\label{fig:persillust}
\vspace{0.25truecm}
\begin{subfigure}[b]{0.32\textwidth}
\includegraphics[width=\textwidth]{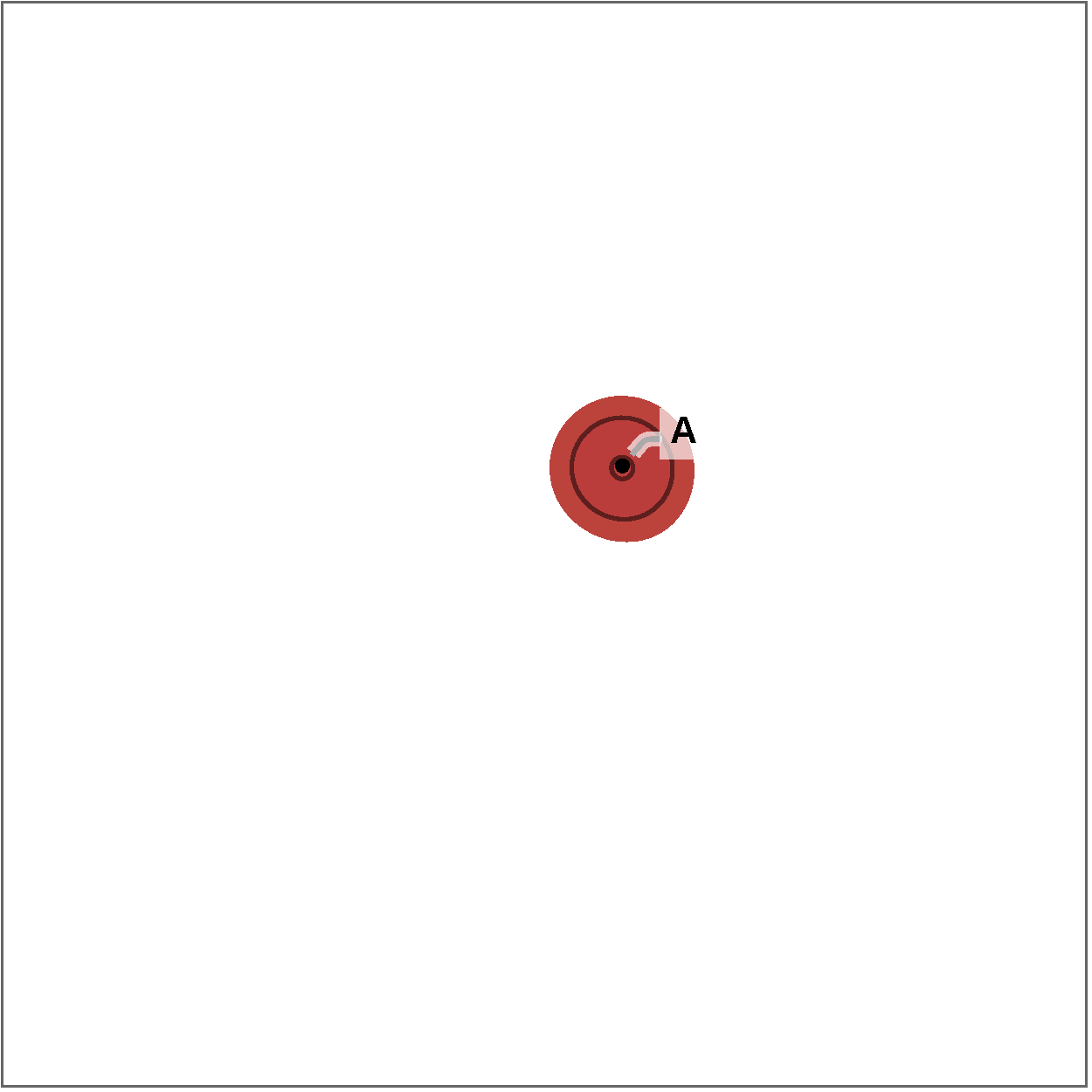}
\end{subfigure}
\begin{subfigure}[b]{0.32\textwidth}
\includegraphics[width=\textwidth]{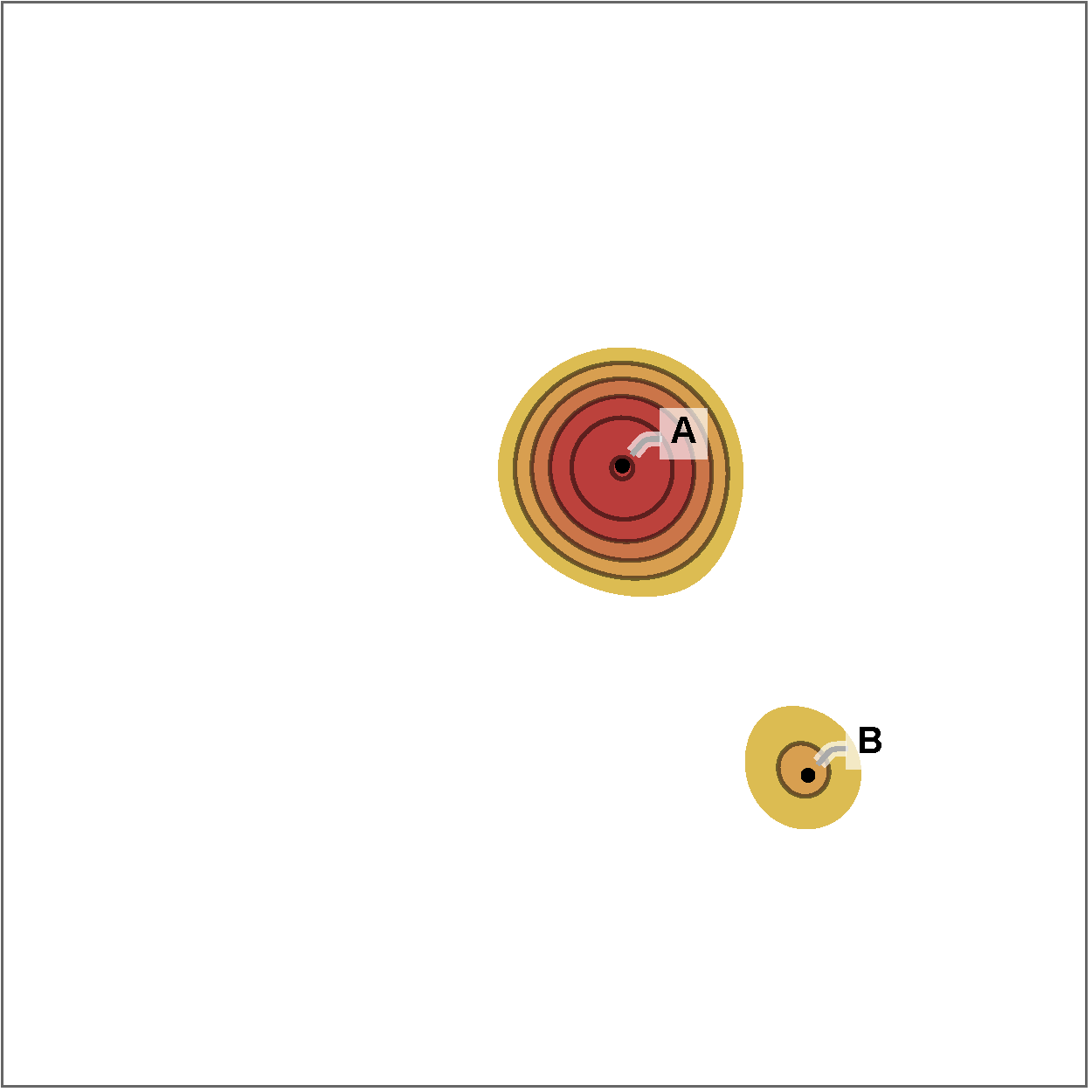}
\end{subfigure}
\begin{subfigure}[b]{0.32\textwidth}
\includegraphics[width=\textwidth]{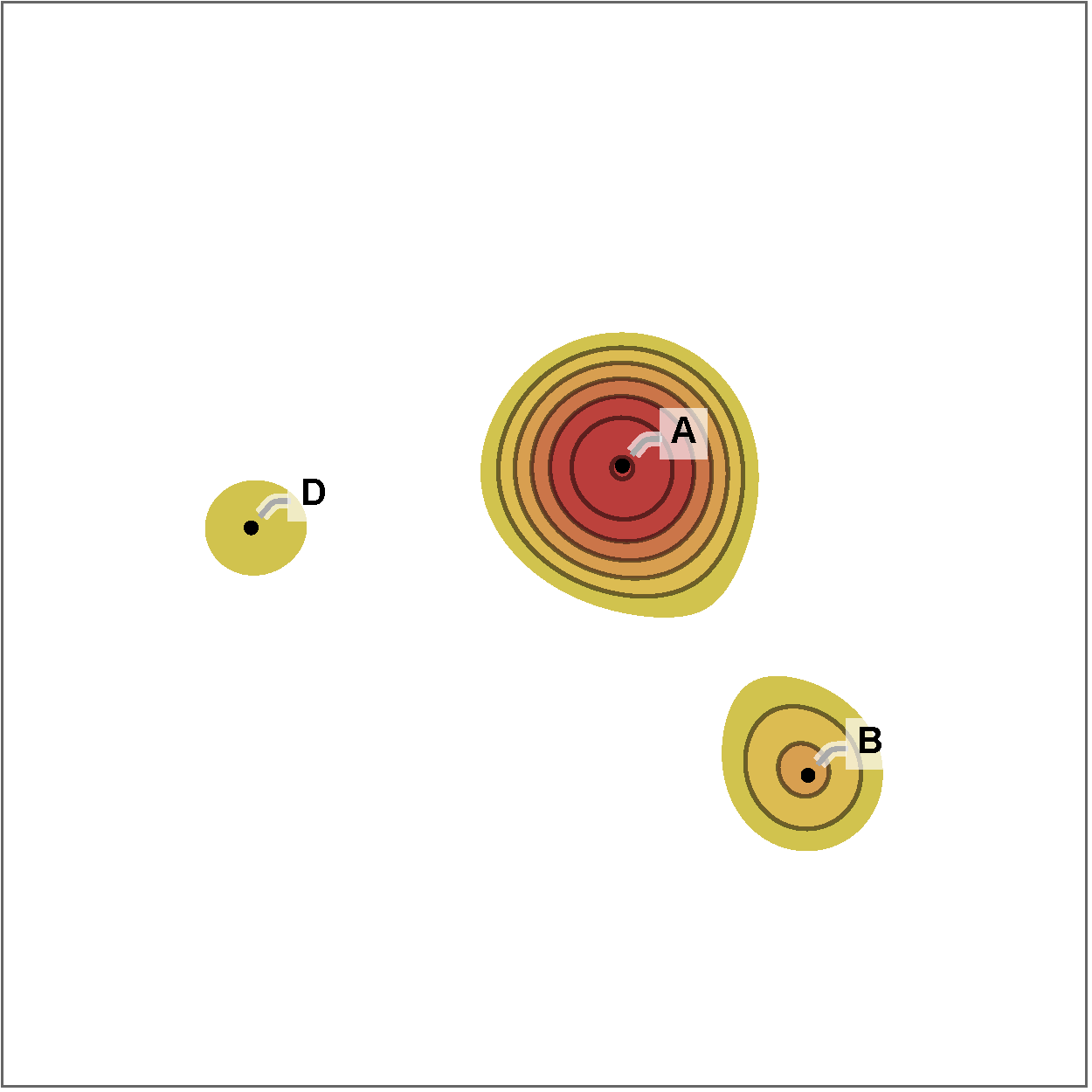}
\end{subfigure}\\
\begin{subfigure}[b]{0.32\textwidth}
\includegraphics[width=\textwidth]{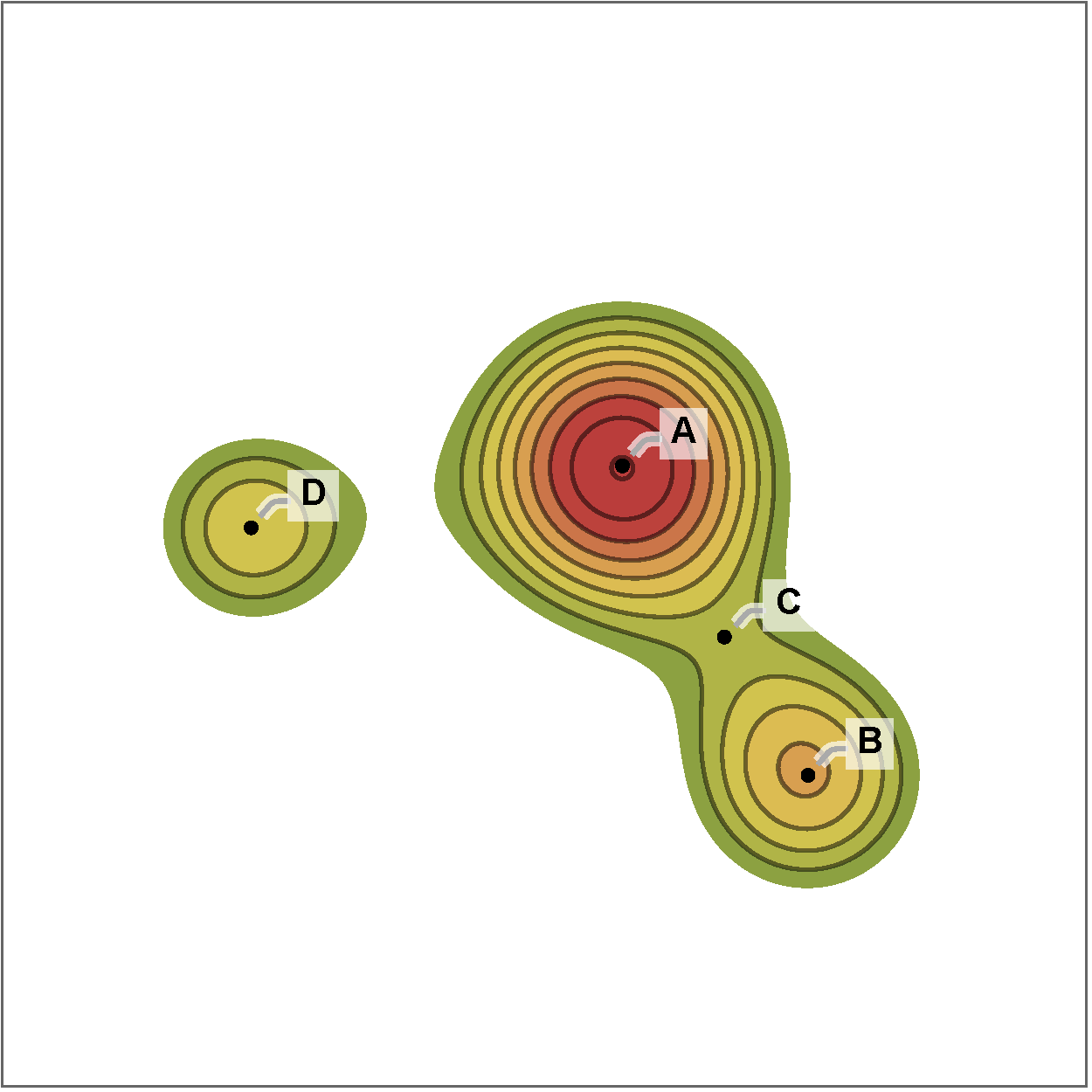}
\end{subfigure}
\begin{subfigure}[b]{0.32\textwidth}
\includegraphics[width=\textwidth]{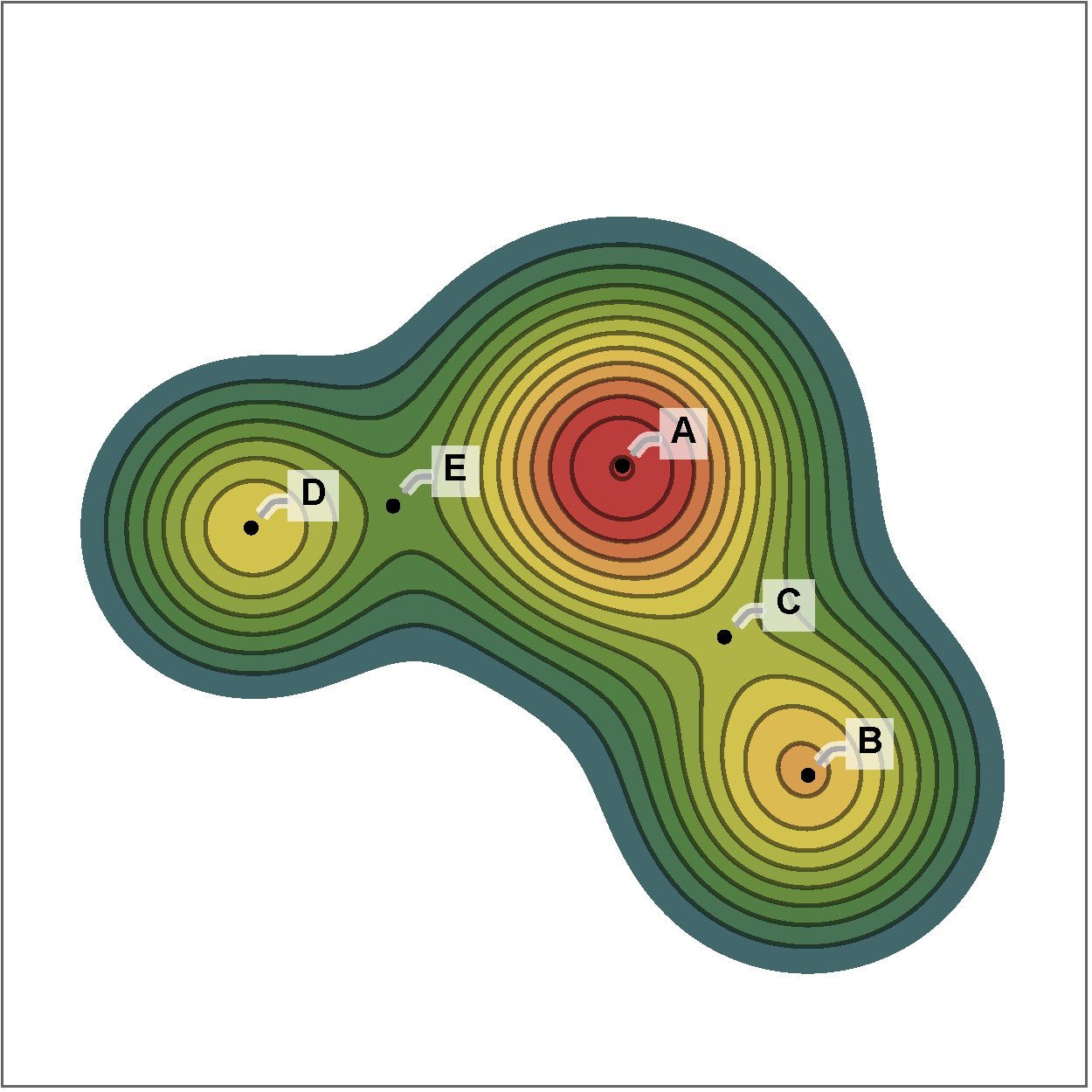}
\end{subfigure}
\begin{subfigure}[b]{0.32\textwidth}
\includegraphics[width=\textwidth]{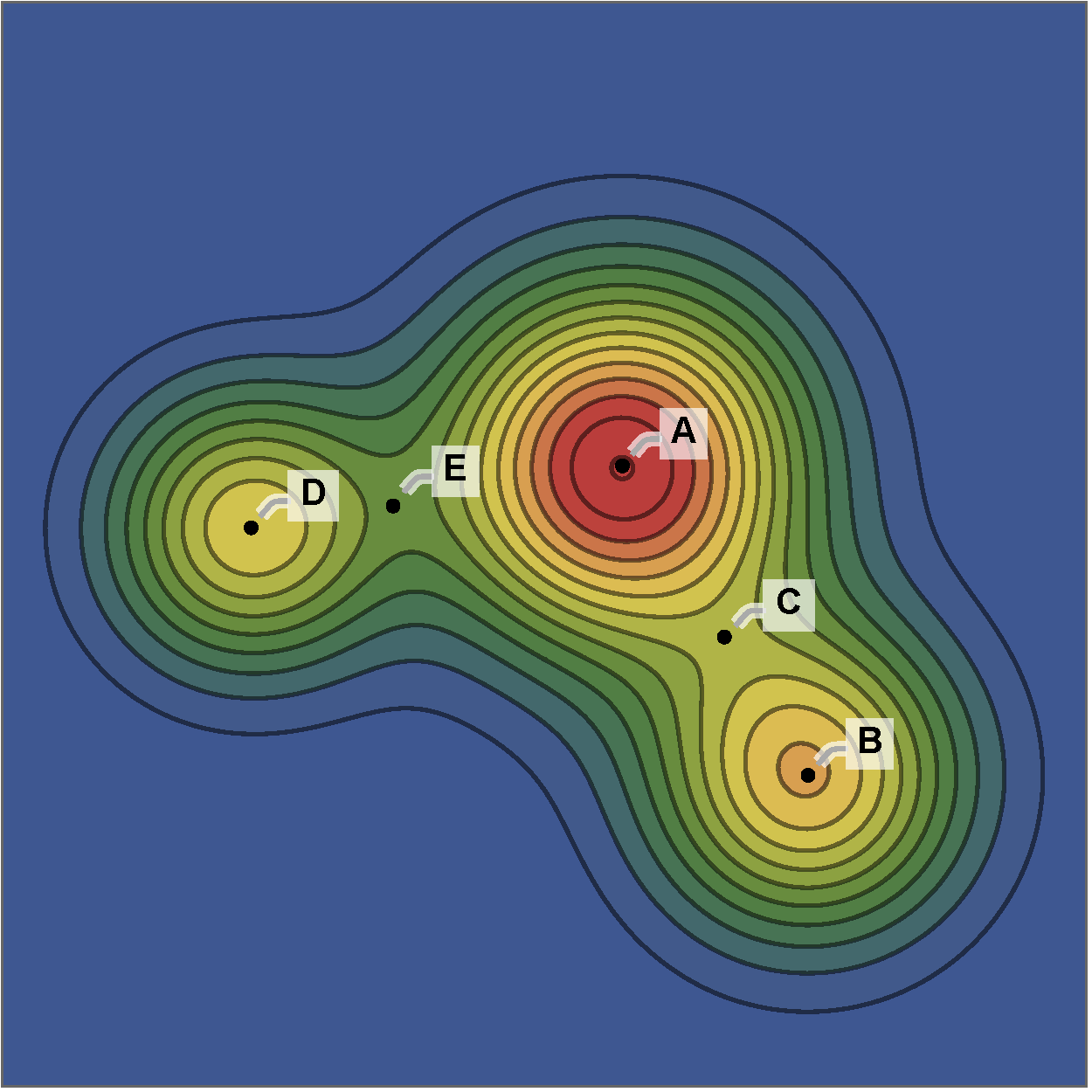}
\end{subfigure}
\end{figure}

\noindent For a visual impression, this expression for $\chi(\nu)$ is shown in figure~\ref{fig:chi}. With respect to
eqn.~\ref{eqn:chigauss}, it is of importance to realize that the analytical expression implicitly assumes the manifold $M$ to be
a smooth, closed, manifold in 2-D Euclidian space $\mathbb{R}^2$. In nearly all practical -- observational -- circumstances this is not true.
Any irregularities, convoluted boundaries, systematic biases, and incompleteness issues will lead to a more complex expression and a
non-symmetric Euler characteristic. The correct expression in those circumstances may be inferred from the {\it Gaussian kinematic formula} or
GKF \citep[see][]{Adler:2009,Adler:2010,Pranav:2019a}. The number density of minima, saddles, and maxima, and the Euler characteristic are
also somewhat modified for a Gaussian random field on the two-sphere, $\mathbb{S}^2$, for which we refer to \cite{Pogosyan:2016}.

\subsection{Persistence Diagrams}
\indent While Betti numbers provide a global characterization of the topology of a manifold, a richer description emerges when we assess the detailed changes in the topological structure of a manifold as the field threshold
changes. 

\begin{figure}
  \centering
  \vspace{0.5cm}
\includegraphics[width=0.65 \textwidth]{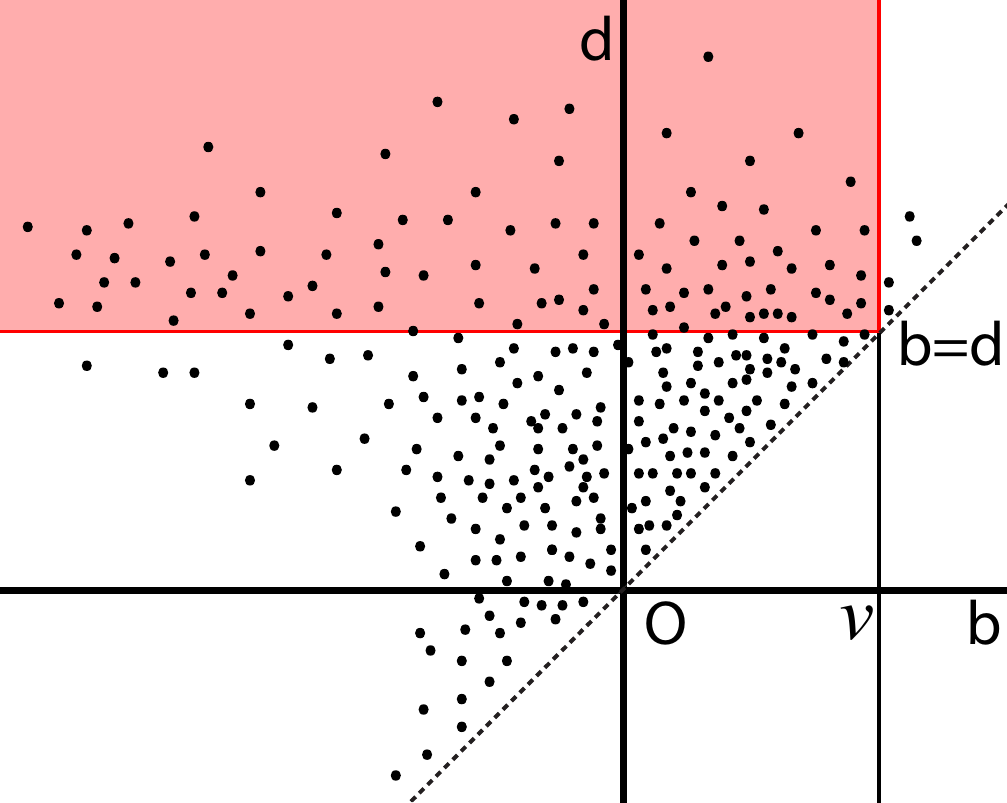}
\vspace{0.5cm}
\caption{Persistence diagram $\Pi_i$, schematic. The diagram plots the birth (b) versus death (d) of a topological feature. The points indicate the
  birth-death (b,d) values for the topological features of a manifold $M$. For a specific superlevel set filtration, $M(\nu)$, the red region indicates
  the birth-death events that mark its topological structure. The number of events in this region yields the corresponding Betti number of $M(\nu)$.}
\label{fig:Pers}
\end{figure}

Complementary to the Betti numbers of a manifold, the structure of the features in the superlevel sets $M(\nu)$ for different threshold values $\nu$ forms a second signature of a random field. As mentioned above, $M(\infty)=\emptyset$. As $\nu$ decreases, connected features enter the superlevel set filtration. If a connected component of $M$ enters the superlevel set filtration at a threshold value $\nu=a$, we define the ($0$-dimensional) feature to be born at $a$.  If the superlevel set $M(\nu)$ consists of at least two connected ($0$-dimensional) components, lowering $\nu$ can lead to the merger of two disconnected components. We define the youngest component to die at $b$, when two features merge at some threshold value $\nu=b$. For example, let two features be born at $a_1=7$ and $a_2=5$ respectively. If these features merge at $b=0$ the second component dies because it is $2$ younger. Similar behaviour can be observed for $1$-dimensional holes.

\indent This process is illustrated in figure \ref{fig:PersExp1}. Initially, at threshold $\nu=\infty$, the superlevel set is the empty set. As the threshold decreases, the maximum $A$ is the first topological feature to appear in the superlevel set.  As the threshold $\nu$ is lowered further, first $B$ appears. $B$ will then merge with $A$ as the saddle point $C$ appears. At this point, the topological feature $B$ dies. We say that the topological feature $A$ is more persistent than $B$. Analogously, as the threshold decreases further, the maximum $D$ is born and subsequently killed by $A$ as the saddle point $E$ appears in the superlevel set filtration. The resulting persistence diagram sorts the topological features in the superlevel sets (or extrema) by their persistence.

\indent As $\nu$ runs from $\infty$ to $-\infty$ all $i$-dimensional topological features
\footnote{A feature is a specified geometric object, while a topological feature is the set of all features which are topologically identical.} 
are born at some threshold $a$ and die at some lower threshold value $b$. Hence each $i$-dimensional feature can be labeled by its birth and death $(a,b)$. 
This pair is called an $i$-dimensional event. The collection of all $i$-dimensional events of a realization of a random field form the $i$-dimensional 
persistence diagram \citep{Edelsbrunner:2009}. The shape of the point cloud, of which the persistence diagram consists, forms a second signature of a 
random field (see figure \ref{fig:Pers}), which has been the subject of several studies including \cite{Adler:2010arXiv1003.1001A, Adler:2013}. A random field over $\mathbb{R}^2$ generically consists of infinitely many connected components and holes. We will, therefore, normalize to the area.

\indent The Betti numbers and persistence diagrams are intimately related. The Betti numbers can be computed from the persistence diagrams (see 
fig.~\ref{fig:Pers}), from which we may immediately infer that they represent a compression of topological information. The Betti numbers at threshold 
$\nu$ are the number of (living) connected components or holes in the superlevel set $M(\nu)$. These components were born before or at $\nu$ and die at a lower threshold. The Betti number $\beta_i(\nu)$ can therefore be obtained by summing over all events $(a,b)$ with $a\geq \nu$ and $b \leq \nu$. In figure \ref{fig:Pers} we see the region over which one should sum to obtain the corresponding Betti number at a specific threshold $\nu$.


\section{The Graph Formalism \& Betti Number Statistics}
\label{sec:graphform}
One of the central tenets of Morse theory is the fact that the full topological structure of a manifold $M(\nu)$ is contained in the
identity, spatial distribution, and connectivity of the critical points of the manifold. Because there are a finite number
of critical points in a volume, the implication is that the full topological structure can be condensed into a discrete
representation of the manifold that properly reflects the spatial singularity distribution. Accordingly, one of the principal
concepts in Morse theory is the {\it Morse-Smale complex} \citep{Morse:1925, Milnor:1963, Edelsbrunner:2009}. It transcribes the topological structure of a
manifold into a spatial tessellation of cells centering around the minima and maxima, in which the edges that define the
cell's perimeter encapsulate the connectivity structure of the manifold. 

\bigskip
\noindent The Morse-Smale complex allows the prospect of developing a formalism for inferring integral expressions for the
expectation values of a manifold's Betti numbers $\beta_i$, and for additional and more intricate homology measures such
as the corresponding persistence diagrams. Here we introduce the {\it Graph Formalism} and describe the steps that lead us
towards the formulation of accurate analytical expressions for Betti numbers. At the core of the formalism is the step in
which the Morse-Smale complex is condensed into a corresponding simplicial complex, whose edges -- 1-faces -- represent the
connections between the various field singularities.

Assessing the evolving network of connecting simplicial edges as we proceed through the manifold filtrations $M(\nu)$ as a function of threshold $\nu$, we witness the unfolding of a network tree. In mathematics, these are known as
graphs \citep{Tutte:2001}.  They establish a profound link between the manifold's topology and graph theory. As the
filtration systematically descends along the graph, and edge by edge is added to the simplicial complex, we witness
the formation, absorption, and demise of simplicial elements in the complex. On the basis of this {\it incremental algorithm}
\citep[see][]{Delfinado:1995,Edelsbrunner:2009}, we are enabled to follow the emergence and disappearance of the
manifold's topological features represented by the corresponding simplicial elements.

With the addition of the crucial aspect of the known probability of critical points -- at density level $\nu$ -- in a
(Gaussian) random field, we may subsequently arrive at a probabilistic evaluation of the formation and destruction of
topological features and assess the expected number of topological features at a field level $\nu$. In the following
subsections we describe how our {\it Graph Formalism} yields integral expressions for the Betti numbers $\beta_0$ and
$\beta_1$ in a two-dimensional setting.

\bigskip
\noindent The integral expressions for the expectation value of the Betti numbers involve the (a priori unknown) probability
function $g(\nu)$ for a topological transition at the level $\nu$. For the Gaussian case, we infer a one-parameter
fitting function for $g(\nu)$, leading us to a closed analytical expression for the expectation values of the
Betti numbers. 

\subsection{Morse-Smale Complex}\label{sec:MorseSmale}
Morse theory is the study of the topology of the level sets of a manifold via the critical points of a smooth scalar function that are
defined on it (see sect.~\ref{sec:critpnt}). A major result of Morse theory is the observation that the global shape and topology of
a manifold is defined only by the character, spatial distribution and connectivity of its critical points \citep{Milnor:1963, Edelsbrunner:2009}.
In a cosmological context, this has been invoked in several studies seeking to delineate the spatial patterns in the cosmic matter
distribution \citep{Colombi:2000,Sousbie:2008,Aragon:2010b,Pranav:2017}.

A key concept in Morse theory is the Morse-Smale complex \citep[see \textit{e.g.}][]{Morse:1925, Gyulassy:2007, Gyulassy:2008}. The Morse-Smale complex is
a partition of space defined by the critical points, and their connections via integral lines, that completely encapsulates the topology
of (filtration) manifolds $M(\nu)$. It consists of a spatial tessellation of stable and unstable manifolds which forms a complete representation
of the topological structure of the manifold. 

\begin{figure}
\includegraphics[width=0.495\textwidth]{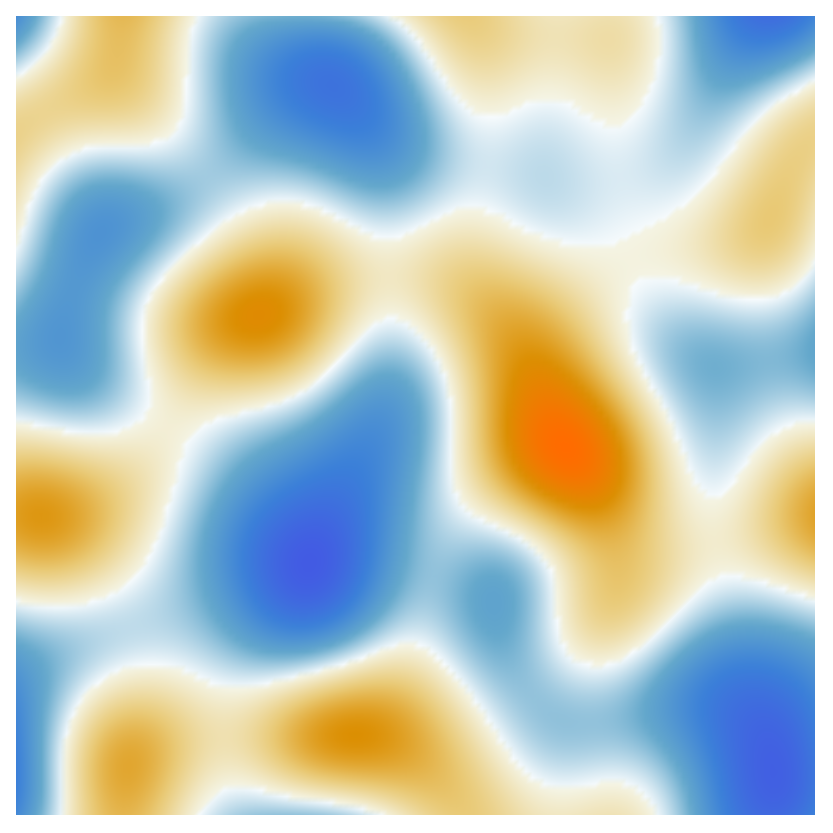}\hfill
\includegraphics[width=0.495\textwidth]{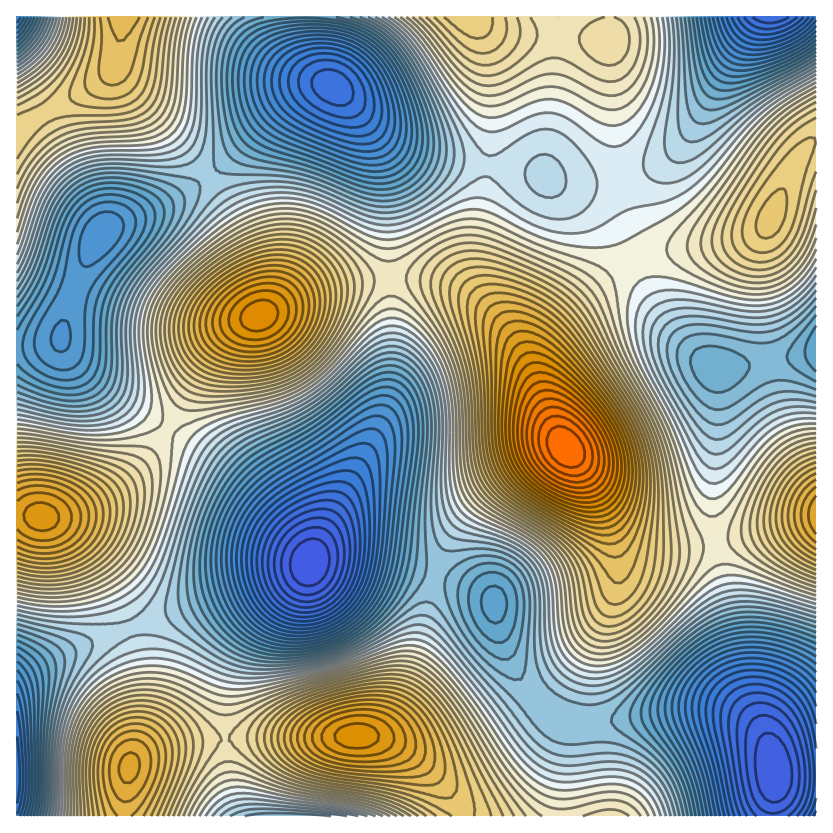}\\
\includegraphics[width=0.495\textwidth]{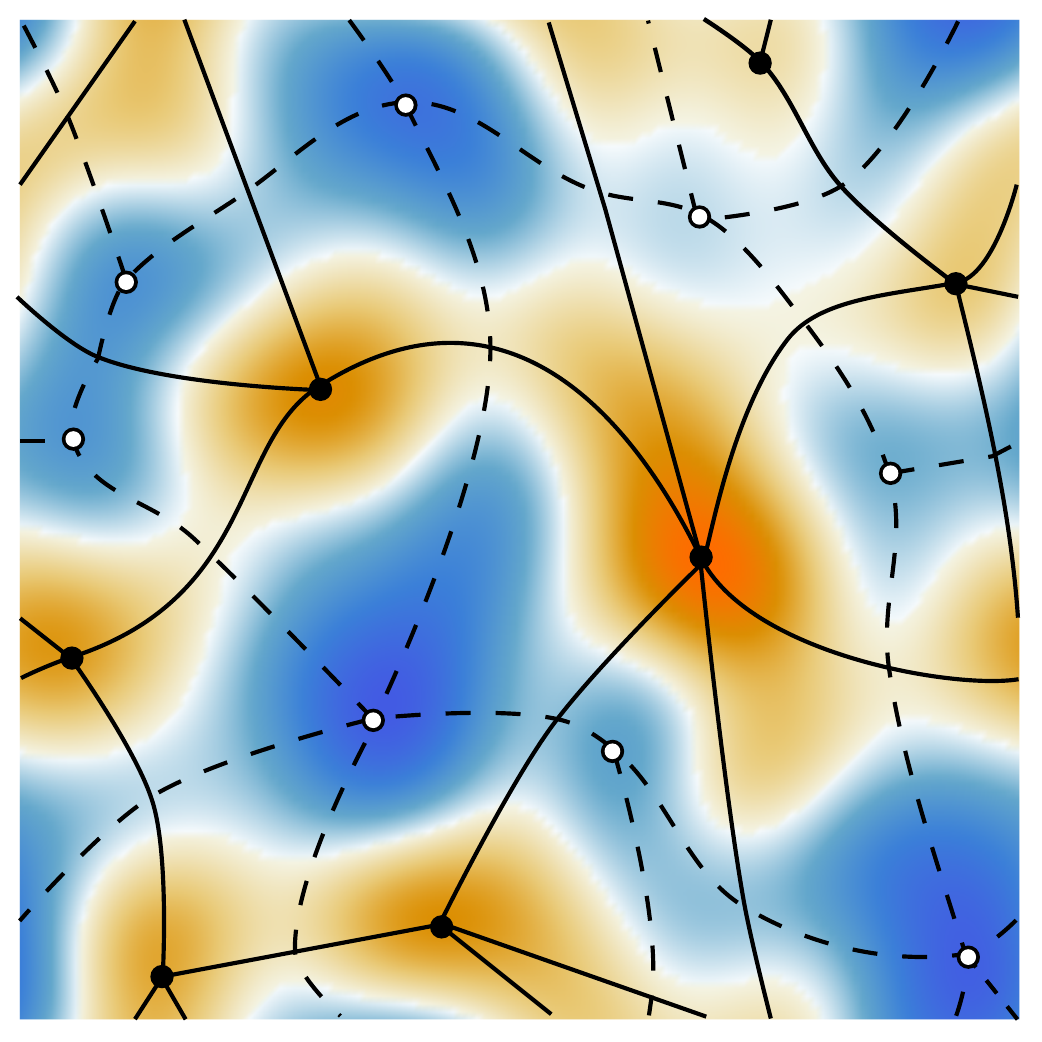}\hfill
\includegraphics[width=0.495\textwidth]{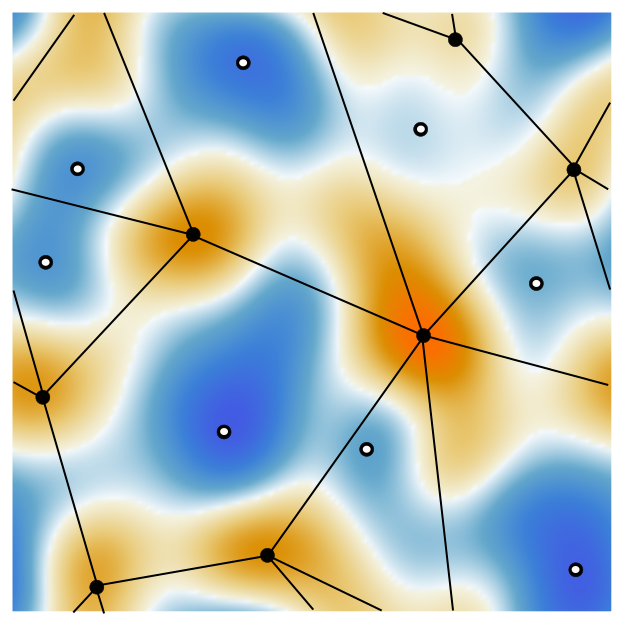}
\caption{Gaussian random field: Singularity structure, Morse-Smale complex and Singularity Graph. The upper left plot is a realization of a Gaussian random field on the unit torus. The upper left figure is an iso-contour plot of the realization in which we can distinguish the minima, saddle points and maxima. The lower left plot is the Morse-Smale complex of the random field. The lower right figure is an impression of the derived tessellation, the Morse-Smale graph, used in our analysis.}
\label{fig:MorseComplex}
\end{figure}

The topological partition and connections between the critical points are established via the {\it integral lines}, which are the paths in the
manifold $M$ whose tangent vectors agree with the gradient of the function $f$ at each point. In other words, the integral lines trace the
gradient flow of the field $f$. Formally, the integral line $\gamma_x:\mathbb{R}\to M$ of a smooth function $f:\mathbb{R}^{2}\to\mathbb{R}$, \footnote{Formally,
  such a function $f$ should be a {\it Morse function}, \textit{i.e.}, a function whose critical points are non-degenerate (\textit{i.e.} its Hessian is non-singular for all
  critical points) and no two critical points have the same function value. Gaussian random fields always obey these conditions.} with
corresponding manifold $M$, is defined as the path satisfying

\begin{align}
\frac{\mathrm{d}}{\mathrm{d}s}\gamma_x(s)=\nabla f(\gamma_x(s)),\ \forall s\in\mathbb{R}\label{IntLine}.
\end{align}

\noindent with boundary condition $\gamma_x(0)=x\in M$. The \textit{origin} and \textit{destination} of an integral line $\gamma$ are defined as

\begin{align}
\text{org}\ \gamma_x=&\lim_{s\to-\infty}\gamma_x(s),\\
\text{dest}\ \gamma_x=&\lim_{s\to \infty}\gamma_x(s).
\end{align}

\noindent Since integral lines always follow the steepest slope of the function $f$, they generically originate and end up at critical points of $f$. 

\bigskip
\noindent Assembling the integral lines with the critical point $e$ as common origin demarcates the region of the manifold $M$ that is known as
{\it unstable} or {\it ascending} manifold. Likewise, one may identify the region that clusters all integral lines for which a critical
point $f$ is the common destination. This delineates the {\it stable} or {\it descending} manifold of the  critical point $f$,

\begin{alignat}{2}
&\mbox{Stable manifold: }\qquad&&S(e)=\{x\in M\ |\ \text{dest}\ \gamma_x=e\}\cup\{e\},\\
&\mbox{Unstable manifold: }\qquad&&U(e)=\{x\in M\ |\ \text{org}\ \gamma_x=e\}\cup\{e\},
\end{alignat}

\medskip
\noindent for all integral lines $\gamma_x$. The decomposition of the manifold $M$ into its stable manifolds is called the Morse complex
of the function $f$. It is a tessellation of $M$ into Morse domains. Likewise, the decomposition of $M$ into its unstable manifolds is
the Morse function of $-f$.

One may obtain a visual impression of the process from figure~\ref{fig:MorseComplex}. It is straightforward to
identify the maxima (red peaks) and minima (blue troughs) in the 2-D Gaussian field realization shown in the top-left frame. In addition
to the presence and location of the peaks and troughs, the contour map in the top righthand frame also reveals nicely the location of the
saddle points. In the bottom left-hand frame of figure~\ref{fig:MorseComplex} we find the Morse complex of stable manifolds delineated by the
dashed black lines originating at the minima in the field. It corresponds to a cellular partition of the manifold into stable Morse domains.
The related Morse complex of unstable manifolds is indicated by the solid lines emanating from the maxima in the field. Furthermore, this defines
a spatial tessellation, consisting of unstable Morse domains. 

\begin{figure}
\includegraphics[width=0.32\textwidth]{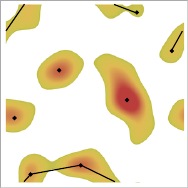}
\includegraphics[width=0.32\textwidth]{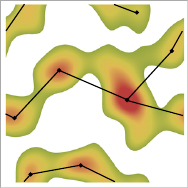}
\includegraphics[width=0.32\textwidth]{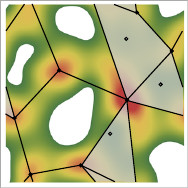}
\caption{Morse-Smale Graph. The 3 frames show the Morse-Smale graph of a manifold for three different filtrations,
  starting from high to low field value $\nu$. Lefthand panel: at high field values, the graph contains a few isolated vertices
  (marking the location of maxima), which in two cases are connected by edges (marking ridges via saddle points). Central panel: at a medium field value, we notice that all vertices have been connected via edges. This marks the formation of larger islands in the manifold, containing several peaks connected by ridges via saddle points. Righthand panel: several edges have connected into a loop/cycle, surrounding a cavity. Three loops even developed into a face, marking a trough with a density higher than the
  threshold.}
\label{fig:MorseGraph}
\end{figure}

The intersections between the stable and unstable Morse domains yields a further decomposition of the manifold $M$ into cells. These
are the {\it Morse-Smale} cells. The Morse-Smale cells are the connected components of the set $U(e_1)\cap S(e_2)$, for all critical points $e_1,e_2\in M$.
Each of the Morse-Smale cells is the domain sharing a common origin ($e_1$) and a common destination ($e_2$). 
The {\it Morse-Smale complex} is the partition of the manifold $M$ into the Morse-Smale cells\footnote{Note that, in general, compactness of the
  manifold $M$ is required for the origin and destination of the Morse-Smale complex to exist. Due to the large number of critical points in realizations of Gaussian random fields, the origin and destination generically exist for a Gaussian random field $f$.}. It is straightforward
to infer the identity of the Morse-Smale cells from the bottom lefthand panel in figure~\ref{fig:MorseComplex}: the intersection between the
stable and unstable domains results in Morse-Smale cells that have a maximum, minimum and two saddle points as a vertex. For a more elaborate analysis
of Morse-Smale complexes see \citep{Edelsbrunner:2003}.

\subsection{Morse-Smale Graph}\label{sec:MorseSmaleGraph}
To condense the information contained in the Morse-Smale complex into one that is straightforward to process for a homology analysis, we translate the Morse-Smale complex of a manifold $M$ into a simplicial complex. In a sense, it is
a straightforward and natural step, following the circumstance that the Morse-Smale complex is a spatial tessellation
of the manifold, which allows it to simplify into a graph representation that still retains the needed topological
information. We call the resulting simplicial complex the {\it Morse-Smale Graph}. 
By construction, the resulting simplicial complex of the {\it Morse-Smale Graph} preserves the homological information of the
superlevel set filtration.

The conversion involves the transformation of maxima into vertices, of saddle points into edges connecting the associated vertices
and of the minima into faces. In the case the manifold for which we seek to construct the simplicial {\it Morse-Smale Graph}
is the superlevel set filtration $M(\nu)$, the graph only includes the simplices -- \textit{i.e.} the vertices, edges and faces -- for which
the function $f$ at the corresponding critical point $\textbf{x}_0$ assumes a value larger or equal than $\nu$,
i.e. $f(\textbf{x}_0) \geq \nu$ (see figure~\ref{fig:MorseGraph}). Dependent on the level, features around maxima (islands) can get
connected. In this case, two vertices are connected by an edge. If the features are disconnected, they are reduced to
vertices in the graph. In other words, two maxima get connected by an edge if and only if they are connected by a saddle point
in the Morse-Smale complex. Likewise, troughs in the manifold $M(\nu)$ are reduced to loops or cycles when the corresponding
minimum is not yet included in the superlevel set (see appendix \ref{appendix:Homology}).

It should be noted that not all Morse-Smale complexes can be reduced to a unique associated simplicial complex. There exist non-degenerate cases in
which parts of the Morse-Smale complex cannot be converted according to the above rules. However, since these cases are non-generic for the random
fields of interest and since we are analysing the statistical properties of homological features, we do not study these cases here.

\subsection{The Incremental Algorithm: \\ \ \ \ \ \ \ \ Graphic Extraction of Homological Information}
In the following, we outline the procedure for the extraction of homological information, by means of an incremental filtration
algorithm. The incremental algorithm is a procedure to compute the Betti numbers of a simplicial complex by tracking the evolution of
the Betti numbers while incrementally introducing simplices \citep[see][]{Delfinado:1995, Edelsbrunner:2009}.

Given a realization of a random field $f:\mathbb{R}^2 \to \mathbb{R}$, we consider the change in homology of the superlevel set
filtration $M(\nu)$ as a function of $\nu$. Starting from infinite density $\nu=\infty$, we follow the evolving Morse-Smale graph as
a function of descending density threshold $\nu$.  Given that -- at each step -- the graph represents the full topology of the manifold filtration
$M(\nu)$. The gradual unfolding of the Morse-Smale graphs enables us to follow the evolution of its topology in terms of the emergence,
merging and annihilation of the corresponding topological features. In the two-dimensional situation, this concerns the appearance
of an ``island'' around a maximum of the field, their merging into larger island features, and the appearance of holes as
such islands connect into features surrounding an empty interior. 

Morse theory teaches us that the topology can only change by inserting or removing a critical point. In other words, the persistent
homology of the Morse-Smale complex changes if and only if a critical point is introduced to the filtration. The maximum introduces a new
connected feature, while the minimum fills up a hole. In the two-dimensional case, the saddle point -- depending on the position of the
saddle point in the Morse-Smale complex -- either connects two connected components or introduces a $1$-dimensional hole to the structure.
The Morse-Smale complex of a realization $f$ thus completely determines the persistent homology.

At threshold $\nu=\infty$ the superlevel set $M(\infty)$, and corresponding Morse-Smale complex, is empty. The corresponding Betti numbers
$\beta_0(\infty),\beta_1(\infty),\beta_2(\infty)$ vanish. As $\nu$ decreases, the elements of the simplicial complex of $M$ are incrementally
added to the simplicial complex of $M(\nu)$. In this process, the zeroth Betti number $\beta_0$ is increased by one for every vertex introduced
to the simplicial complex. Similarly, $\beta_1$ is decreased by one for every face introduced to the simplicial complex. The introduction of an edge -- corresponding to a saddle point -- either connects two previously disconnected parts, decreasing $\beta_0$ by one or introduces a cycle to the simplicial complex, increasing $\beta_1$ by one. The incremental algorithm is summarized in table \ref{tab:Betti}. The second Betti number $\beta_2$ is included, as the addition of the last face of the graph formally increases $\beta_2$ from $0$ to $1$.\\

\noindent Based on the steps of the incremental procedure outlined in table \ref{tab:Betti}, we may subsequently infer the following integral relations
for $\beta_0$, $\beta_1$ and $\beta_2$:

\begin{align}
  \beta_0(\nu) = &\int_{\nu}^\infty[ \mathcal{N}_2(\mu)-(1-g(\mu))\mathcal{N}_1(\mu)]\mathrm{d}\mu\label{eq:beta0}\text{,}  \\
\beta_1(\nu) = &\int_{\nu}^\infty[ g(\mu)\mathcal{N}_1(\mu)-\mathcal{N}_0(\mu)]\mathrm{d}\mu \label{eq:beta1} \text{,} \\
\beta_2(\nu) =& 0\,,
\end{align}

\medskip
\noindent with $\mathcal{N}_0,\mathcal{N}_1,\mathcal{N}_2:\mathbb{R}\to\mathbb{R}$ the density of minima, saddle points and maxima at threshold $\nu$.
In these integral expressions, the function $g(\nu)$ is the probability that the introduction of a saddle point at threshold $\nu$ creates a cycle
increasing $\beta_1$, 

\begin{align}
g(\nu)  = P\left[\text{At }\nu, \beta_1 \mapsto \beta_1 + 1| \text{ At } \nu \text{ a saddle point is added to the simplicial complex}\right]\,,\label{eq:gdef}
\end{align}

\medskip
\noindent with $g:[-\infty,\infty] \to [0,1]$. Inversely, the function $1-g(\nu)$ denotes the probability that a saddle point connects two disconnected features
and decreases $\beta_0$ by one, \textit{i.e.},

\begin{align}
1-g(\nu)  = P\left[\text{At }\nu, \beta_0 \mapsto \beta_1 - 1| \text{ At } \nu \text{ a saddle point is added to the simplicial complex}\right]\,.
\end{align}

\medskip
\noindent Following the above, the problem of computing the Betti numbers $\beta_0,\beta_1$ of a random field has been reduced to
the determination of the probability $g$. Note that this formulation applies to arbitrary two-dimensional random fields, \textit{i.e.}, it does
not presume the Gaussianity condition.

\subsection{Approximating the Probability $g$}
In order to deduce an accurate approximation for the probability function $g(\nu)$, we investigate its asymptotic behaviour in the
limits $\nu \to \pm \infty$. 

In the limit $\nu = \infty$, the superlevel set filtration $M(\infty)$ is empty and the Betti numbers vanish. As the threshold $\nu$ is decreased,
we first see the addition of isolated islands. These are $0$-dimensional vertices in the corresponding simplicial complex.
These vertices dominate over the edges and faces, as at high function values, as maxima are more numerous
than saddle points and minima (see figure \ref{fig:criticalPoint}). At high $\nu$, saddle points are therefore more likely to connect disjoint vertices than to form cycles. This
translates into a simplicial complex that consists of many isolated vertices and relatively few clusters.

As we proceed to medium levels of $\nu$, we encounter the reverse situation (see figure \ref{fig:criticalPoint}). The introduction of a saddle point is likely to form a cycle since the
simplicial complex consists of many clusters and relatively few connected components. As a consequence, as we decrease $\nu$ starting from
$\nu = \infty$, the Betti number $\beta_0$ is first to increase while the Betti number $\beta_1$ remains approximately constant. At low $\nu$, we
reach a phase in which $\beta_1$ decreases whereas $\beta_0$ is approximately constant. At $\nu=-\infty$, the superlevel set filtration $M(\nu)=M$ and
the Betti numbers are determined by the homological structure of $M$. Differentiating equations \eqref{eq:beta0} and \eqref{eq:beta1} in these two
regimes results in the asymptotic behaviour

\begin{align}
g(\nu) &\sim g_{+}(\nu)\,\equiv\,\frac{\mathcal{N}_0(\nu)}{\mathcal{N}_1(\nu)}     &&\text{as \ $\nu\to \infty$}\,, \label{eq:asympt1}\\
g(\nu) &\sim g_{-}(\nu)\,\equiv\,1-\frac{\mathcal{N}_2(\nu)}{\mathcal{N}_1(\nu)}     &&\text{as \ $\nu \to -\infty$}\,.\label{eq:asympt2}
\end{align}

\medskip
\noindent These asymptomatics apply to general two-dimensional random fields.
\begin{table}
\begin{center}
  \begin{tabular}{ |p{4cm}||c||c|c|c||p{2cm}| }
    \hline
    & & & & & \\
Critical point & index & $\beta_0$ & $\beta_1$ & $\beta_2$& addition of \\ 
    & & & & & \\
\hline
    & & & & & \\
Minimum & 0 & &$\downarrow$&$\uparrow$&face\\
    & & & & & \\
Saddle point & 1 & $\downarrow$&$\uparrow$&&edge\\ 
    & & & & & \\
Maximum & 2 &$\uparrow$ &&& vertex \\ 
    & & & & & \\
\hline
  \end{tabular}
\end{center}
  \caption{Betti number trend as a function of superlevel filtration threshold: decrease ($\downarrow$)/increase ($\uparrow$) with decreasing threshold in the incremental algorithm.}
\label{tab:Betti}
\end{table}

\medskip
Subsequently restricting ourselves to a Gaussian random field, it is straightforward to infer that the derivatives of the Betti numbers at $\nu=0$
with respect to the threshold are related by
\begin{align}
\frac{\mathrm{d}\beta_0}{\mathrm{d}\nu}\bigg|_{\nu=0}=-\frac{\mathrm{d}\beta_1}{\mathrm{d}\nu}\bigg|_{\nu=0}.
\end{align}
From the integral expressions \eqref{eq:beta0} and \eqref{eq:beta1}, we then easily infer a relation between the
number densities $\mathcal{N}_0(0)$, $\mathcal{N}_1(0)$, $\mathcal{N}_2(0)$ of maxima, saddles and minima at $\nu=0$,
\begin{align}
-\mathcal{N}_2(0)+\left[1-g(0)\right]\mathcal{N}_1(0)&=g(0)\mathcal{N}_1(0)-\mathcal{N}_0(0)\,.\label{eq:condition}
\end{align}
\noindent In addition, the statistical reflection invariance of Gaussian random fields, $f \mapsto -f$, implies a statistical
symmetry between the corresponding Betti numbers, $\beta_0(\nu) = \beta_1(-\nu)$, and number densities, $\mathcal{N}_0(0)=\mathcal{N}_2(0)$.
From this we obtain the additional constraint 
\begin{align}
g(0)=\frac{1}{2}\,.\label{eq:constraint}
\end{align}

\begin{figure}
\centering
\includegraphics[width=0.85\textwidth]{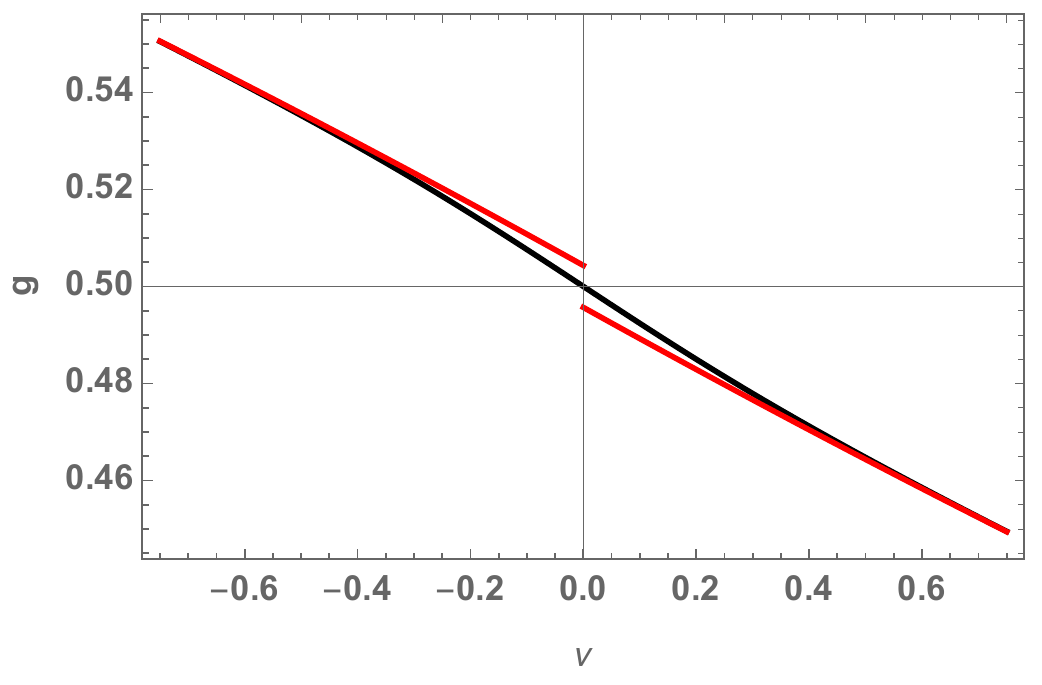}
\vskip -0.25truecm
\caption{Probability distribution $g$: comparison between function $g(\nu)$ (black) and the 2 fitting relations (red) for
  $g_{-}(\nu)$ for $\nu < 0$ and $g_{+}(\nu)$ for $\nu > 0$. For their expressions see eqns. \eqref{eq:asympt1}, \eqref{eq:asympt2} and \eqref{eq:gfit}.}
\label{fig:gnu}
\centering
\vspace{0.5cm}
\mbox{\hskip -0.4cm\includegraphics[width=0.84\textwidth]{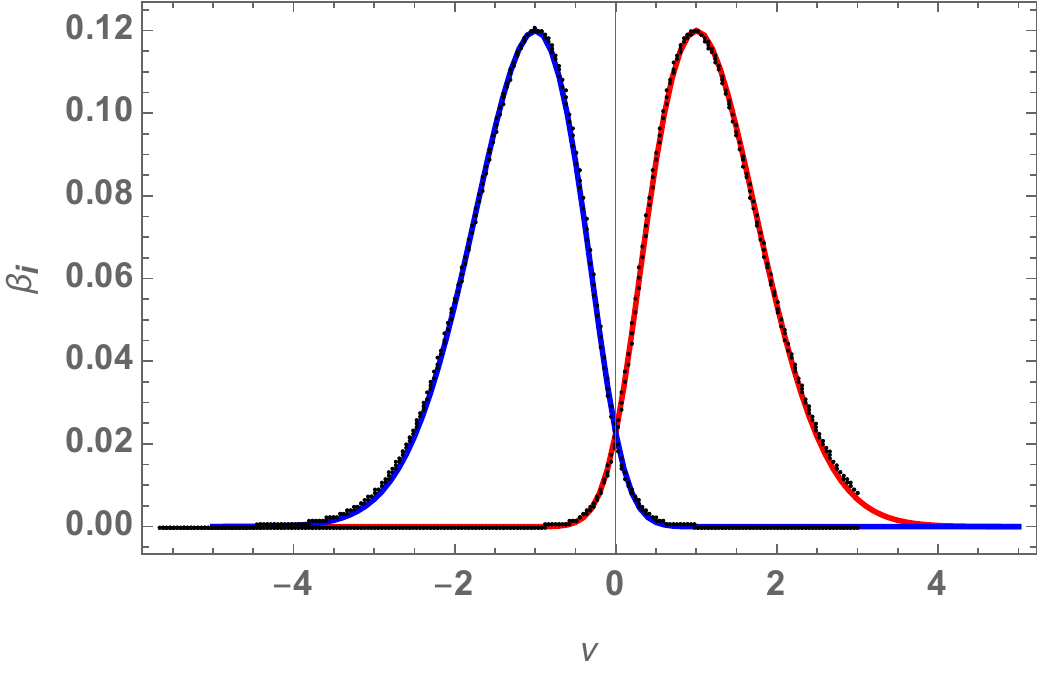}}
\vskip -0.25truecm
\caption{Betti numbers, fitting formula \textit{v.s.} numerical Betti curve. Lines: analytical relation for the Betti curves $\beta_0(\nu)$ (red) and $\beta_1(\nu)$ (blue) for a two-dimensional Gaussian random field (with a LCDM power spectrum), as specified by expressions \eqref{eq:beta0} and \eqref{eq:beta1}. Dots: the corresponding numerically computed Betti numbers for a realization of the same LCDM Gaussian random field. }
\label{fig:Betti}
\end{figure}

\bigskip
\noindent Following the asymptotic relations \eqref{eq:asympt1}, \eqref{eq:asympt2} and constraint \eqref{eq:constraint}, we are therefore
led to propose the following fitting formula for $g(\nu)$, 
\begin{align}
g(\nu)\approx{\displaystyle \frac{g_{+}(\nu)\,e^{\alpha\nu}\,+\,g_{-}(\nu)\,e^{-\alpha\nu}}{e^{\alpha\nu}+e^{-\alpha\nu}}}\,\text{,}
\label{eq:gfit}
\end{align}
\medskip
\noindent which includes the fitting parameter $\alpha$. Note that the proposed fitting function has the asymptotic limits,
\begin{align}
\lim_{\nu \to -\infty} g(\nu) &= 1\,,\nonumber\\
\lim_{\nu \to \infty}g(\nu)&=0\,,
\end{align}
as expected. For the fitting function $g$ and the two asymptotics for a Gaussian random field near $\nu =0$ we refer to figure \ref{fig:gnu}. The two asymptotics $\mathcal{N}_0(\nu)/\mathcal{N}_1(\nu)$ and $1- \mathcal{N}_2(\nu)/\mathcal{N}_1(\nu)$ are very close to the constraint $g(0)=1/2$ at $\nu=0$.
The fitting function $g$ is a small quantitative correction and quickly approaches the asymptotes for non-zero $\nu$. On the other hand, the correction
is significant for the behaviour of the Betti numbers $\beta_i(\nu)$ near $\nu = 0$. For a Gaussian random field, the parameter $\alpha$ should be expressible in terms of the power spectrum $P(k)$. This relation will be the subject of an upcoming study.

Given the fitting function for $g$, equations \eqref{eq:beta0} and \eqref{eq:beta1} form a one-parameter fitting function for the Betti curves $\beta_0(\nu)$ and $\beta_1(\nu)$. In figure \ref{fig:Betti}, we compare the Betti numbers measured for a realization of a Gaussian random field with the
best fitting function $g(\nu)$. Equation \eqref{eq:gfit} turns out to be a good approximation of the probability that a saddle point leads to
a cycle. Also, equations~\eqref{eq:beta0} and \eqref{eq:beta1} capture all the features of the Betti curves. We have attempted several alternative interpolation schemes for equation \eqref{eq:gfit}. However, the Betti numbers predicted by equations \eqref{eq:beta0} and \eqref{eq:beta1} turn out to be very sensitive to the behaviour of $g$ near $\nu$. Many weightings can quickly be ruled out, while the exponential weighting turns out to lead to the most accurate results.

\subsection{Numerical Validation of $g$}
In figure \ref{fig:Betti}, numerically determined Betti numbers and the analytically predicted Betti numbers of a two-dimensional Gaussian random field are shown. The analytic approximation deviates at $\nu=0$ and at larger $|\nu|$. Nonetheless, the approximation appears to capture the qualitative nature of
the Betti numbers. For this plot, we used a fitting parameter $\alpha=2.71$. Empirically, the Betti numbers turn out to be very sensitive to the choice of
the procedure for gluing the two branches of $g(\nu)$ and to the value of the fitting parameter. In turn, this implies that equation \ref{eq:gfit} must be a
good approximation of $g$ and $\alpha$ should be approximately constant.


\section{Statistics of Persistence diagrams of Random Fields}
\label{sec:persistence}
In the previous section, we inferred an (approximate) analytical expression for the expected persistent Betti numbers of a Gaussian random field, in terms of the distribution of critical points. Here we forward integral expressions for the persistence diagrams of two-dimensional Gaussian random fields, following the same graph formalism. 

The $i$-dimensional persistence diagram $\Pi_i$ of the superlevel set filtration $M(\nu)$ of a function $f$ consists of the set of events $(b,d)$ corresponding to the birth at $f = \sigma b$ and death at $f = \sigma d$ of an $i$-dimensional topological feature. The persistence diagram of a random field is a continuous function $\Pi_i:\mathbb{R}^2\to \mathbb{R}$ defined as the expectation value
\begin{align}
\Pi_i(b,d) = \text{E}[\text{number of events } (b,d) \text{ per unit area/volume in a realization}]\,.
\end{align}
Using the graph formalism presented in section \ref{sec:MorseSmaleGraph}, the persistence diagram $\Pi_i$ is directly related to the distribution of and connections between critical points. Following the growth of the Morse graph as we proceed through the field filtration, singularities get absorbed into the graph and establish new graphic connections. As they do so, topological features are born or vanish. The levels at which a feature emerges and at which it disappears defines a (birth,death) point in the persistenc diagram. 
 
\medskip
\noindent The zero-dimensional persistence diagram $\Pi_0$ consists of the (birth, death) pairs that represent the birth of an island and the subsequent merging of this feature with a neighbouring island. The birth of an island corresponds to the appearance of a local maximum at function value $f=\sigma b$ in the Morse graph. The merger with a neighbouring island occurs upon the introduction of a saddle point at function value $f=\sigma d$. The saddle point joins
two disconnected components. Evidently, it is the birth \textit{c.q.} death of an island feature that contributes to the evolution of the zeroth persistent
Betti number $\beta_0(\nu)$. 

The (birth, death) pairs in the one-dimensional persistence diagram $\Pi_1$ represents the formation of a loop and the subsequent filling up of the loop.
The creation of the loop corresponds to the introduction of a saddle point at function value $f=\sigma b$, while the loop fills up and disappears
due to the introduction of a minimum at function value $f=\sigma d$. In the case of $\Pi_1$, a saddle point connects two features which were already
connected. It is self-evident that this phenomenon relates to the evolution of the first persistence Betti number $\beta_1(\nu)$.\\

\bigskip
Following the observations above, we may formulate the corresponding expressions for the expected persistence diagrams $\Pi_i$ of a two-dimensional
random field in terms of the statistics of the critical points. As a function of the birth and death field levels $(b,d)$, these are the integral expressions
\begin{align}
\Pi_0(b,d)&=\int\limits_{0}^{\infty} G_0(b,d,r)p_0(b,d,r)\mathrm{d}r\,,\\
\Pi_1(b,d)&=\int\limits_{0}^{\infty} G_1(b,d,r) p_1(b,d,r)\mathrm{d}r\,, 
\end{align}
in which $p_0(a,b,r)$ is the number density of (maximum, saddle point)-pairs whose mutual distance is $r=\|\textbf{x}_m - \textbf{x}_s\|$,
where the maximum is located at $\textbf{r}_m$, at function value $f(\textbf{r}_m)=\sigma b$, and the saddle point at position $\textbf{r}_s$, at function value $f(\textbf{r}_s)=\sigma d$. Likewise, $p_1(b,d,r)$ is the number density of (saddle point, minimum)-pairs with $r$ the distance between
the saddle point, at function value $b$, and the minimum at function value $d$.

The number densities $p_1$ and $p_0$ are natural extensions of the number density of the field singularities, $\mathcal{N}_0$, $\mathcal{N}_1$ and
$\mathcal{N}_2$, whereby they incorporate the spatial correlations between the critical points. Using Rice's formula \cite{Rice:1944, Longuet-Higgins:1957, Bardeen:1986, Adler:1981, Adler:2009} we can express
$p_0$ and $p_1$ as the four-dimensional integral
\begin{align}
p_i(b,d,r)= \iiiint &P(b,f_k(0)=0,J_k(0),d,f_k(r)=0,J_k(r))\nonumber\\
&\times\big (|J_1(0)^2-J_2(0)|+|J_1(r)^2-J_2(r)|\big) \mathrm{d}J_1(0)\mathrm{d}J_2(0)\mathrm{d}J_1(r)\mathrm{d}J_2(r)\,,
\end{align}
where the integral for $i=0$ and $i=1$ ranges over $(J_1(0),J_2(0),J_1(r),J_2(r))\in(-\infty,0)\times(0,J_1(0)^2)\times(-\infty,\infty)\times(J_1(r)^2,\infty)$ and $(J_1(0),J_2(0),J_1(r),J_2(r))\in(-\infty,\infty)\times(J_1(0)^2,\infty)\times(0,\infty)\times(0,J_1(r)^2)$.

In the intergral expression for persistence $\Pi_0(b,d)$, the function $G_0(b,d,r)$ is the probability that the introduction of the saddle point
with function value $f = \sigma d$ at location $\textbf{r}_s$ induces the death of the island corresponding to the maximum with function value $f = \sigma b$
at location $\textbf{r}_m$. Likewise, the function $G_1$ is the analogous probability that the saddle point in the (saddle point, minimum)-pair induces
the introduction of a loop at function value $f=\sigma b$ which is killed by the minimum at function value $f= \sigma d$. The two functions $G_0$ and $G_1$
are natural generalizations of the probability function $g$ (equation \eqref{eq:gdef}). The relation can be made explicit by integrating over the birth $b$ and death
$d$ variables
\begin{align}
g(\nu)=\int\limits_0^\infty\int\limits_{-\infty}^\nu\int\limits_\nu^\infty G_0(b,d,r) \mathrm{d}b\,\mathrm{d}\,d\mathrm{d}r\,.
\end{align}
The integral over the separation $r$ traces over the different configurations which lead to the same birth-death event $(b,d)$ in the
persistence diagram. \\

The statistical symmetry of Gaussian random fields with respect to reflection through the mean $\mu = 0$ ensures that the persistence diagrams $\Pi_0,\Pi_1$ are statistically symmetric under reflections through the line $b = -d$, \textit{i.e.}, $\Pi_0(b,d) = \Pi_1(d,b)$. In addition, from the Morse-Smale complex we deduce that the probability $G_0$ is related to the probability that in a realization of a random field $f$, the points $A=(0,0)$ and $B=(r,0)$, with $A$ a maximum and $B$ a minimum of $f$ are connected by an integral line $\gamma$ between $A$ and $B$ defined by
\begin{align}
\gamma(t)\in M\text{ and }\frac{\mathrm{d}\gamma(t)}{\mathrm{d}t}=\nabla f(\gamma(t)) \ \forall t\in \mathbb{R}\,.
\end{align}
For Gaussian random fields, we expect this probability to be expressible in terms of a Gaussian path integral \citep{Feynman:1965}. Such an expression would, besides predicting the persistence diagrams $\Pi_i$, yield an expression for the fitting parameter $\alpha$ in terms of the power spectrum. Obtaining an approximation for $Q_0$ and $g$ using path integral methods is the subject of an upcoming paper \citep{Feldbrugge:2020}.

\begin{figure*}
\centering
\includegraphics[width=\textwidth]{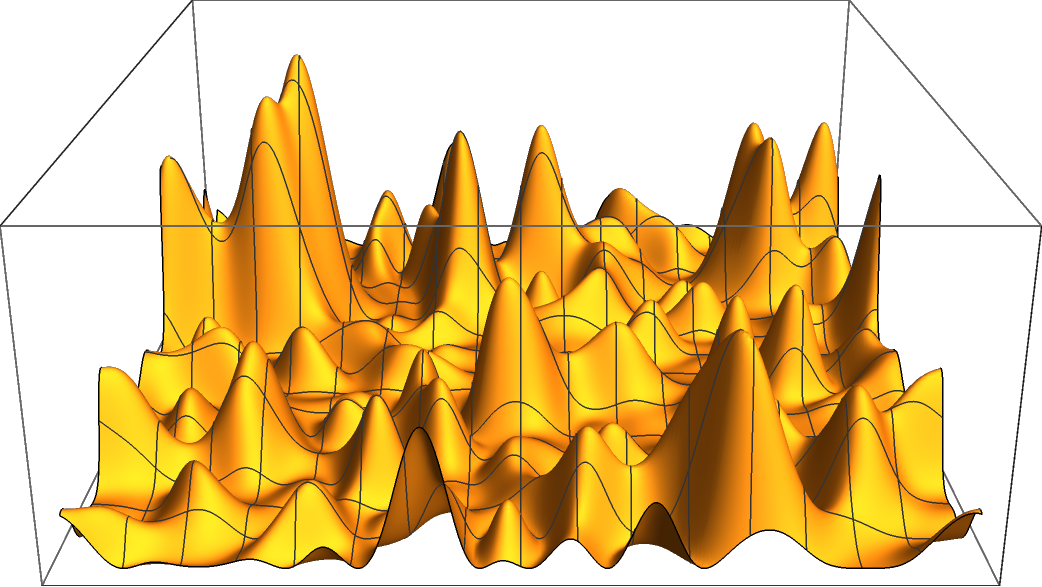}
\caption{A highly local non-Gaussian random field of the local template with non-Gaussianity parameter $f_{NL}=1$. From this field one can clearly observe an asymmetry in the statistics of maxima and minima.}
\label{fig:NonGaus}
\end{figure*}

\section{Non-Gaussian Random Fields: Persistence \& Betti numbers}
\label{sec:nongauss}
In the previous section, we analysed the persistent Betti numbers of the superlevel set filtration $M(\nu)$ of a random field $f:\mathbb{R}^2\to\mathbb{R}$, and derived the corresponding fitting formulas for the Gaussian case. The analysis can be extended naturally to random fields on more general two-dimensional manifolds. When applied to random fields on the two-dimensional sphere $\mathbb{S}^2$, the Gaussian persistent Betti numbers can be compared to the persistent homology of the CMB temperature perturbations $\delta T$ to look for deviations from Gaussianity \citep{Planck:2016, Planck:2019, Pranav:2019b}. In this section, we numerically investigate the sensitivity of the persistent Betti numbers and persistence diagrams on non-Gaussianities in random fields on $\mathbb{R}^2$. An investigation of the behaviour of persistent homology of random fields on the sphere $\mathbb{S}^2$ is the subject of an upcoming paper \citep{Feldbrugge:2020b}.

\begin{figure}
\centering
\begin{subfigure}[b]{0.325\textwidth}
\caption{$\Pi_0$,\,\,$f_{NL}=0$}
\includegraphics[width=\textwidth]{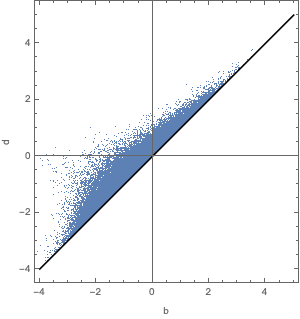}
\label{fig:nl0}
\end{subfigure}
\begin{subfigure}[b]{0.325\textwidth}
\caption{$\Pi_0$,\,\,$f_{NL}=0.05$}
\includegraphics[width=\textwidth]{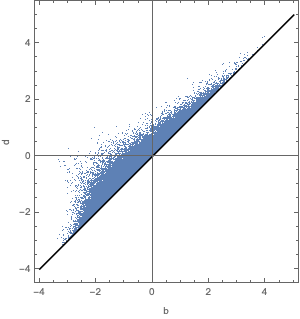}
\label{fig:nl005}
\end{subfigure}
\begin{subfigure}[b]{0.325\textwidth}
\caption{$\Pi_0$,\,\,$f_{NL}=0.1$}
\includegraphics[width=\textwidth]{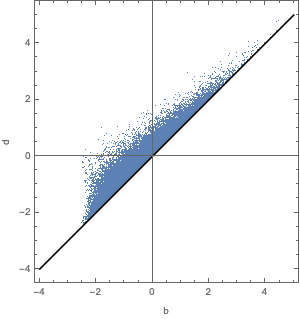}
\label{fig:nl01}
\end{subfigure}\\
\begin{subfigure}[b]{0.325\textwidth}
\mbox{\hskip 0.1cm\includegraphics[width=\textwidth]{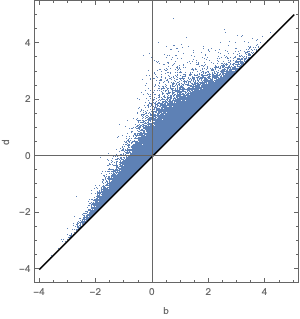}}
\caption{$\Pi_1$,\,\,$f_{NL}=0$}
\label{fig:nl0_2}
\end{subfigure}
\begin{subfigure}[b]{0.325\textwidth}
\mbox{\hskip 0.1cm\includegraphics[width=\textwidth]{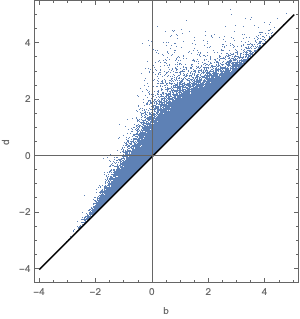}}
\caption{$\Pi_1$,\,\,$f_{NL}=0.05$}
\label{fig:nl005_2}
\end{subfigure}
\begin{subfigure}[b]{0.325\textwidth}
\mbox{\hskip 0.1cm\includegraphics[width=\textwidth]{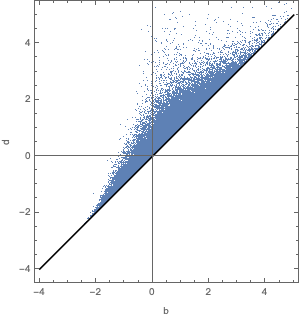}}
\caption{$\Pi_1$,\,\,$f_{NL}=0.1$}
\label{fig:nl01_2}
\end{subfigure}
\hspace*{\fill}
\caption{Persistence Diagrams non-Gaussian random fields. The figure shows the $0$- and $1$-dimensional persistence diagram of a Gaussian and two non-Gaussian random fields. The non-Gaussian fields are of a local nature, characterized by parameter $f_{NL}$. Left:  a Gaussian random field, with LCDM power spectrum (see text). By definition, it has non-Gaussianity parameter $f_{NL}=0$. Centre: a random field with
  local non-Gaussianity parameter $f_{NL}=0.05$. Right: a random field with local non-Gaussianity parameter $f_{NL}=0.1$. The persistence diagrams clearly reveal
  a significant change in shape as a function of $f_{NL}$. }
\label{fig:persBBKS}
\end{figure}

\begin{figure}
\begin{subfigure}[b]{0.32\textwidth}
\caption{$f_{NL}=0$}
\mbox{\hskip 0.2cm\includegraphics[width=\textwidth]{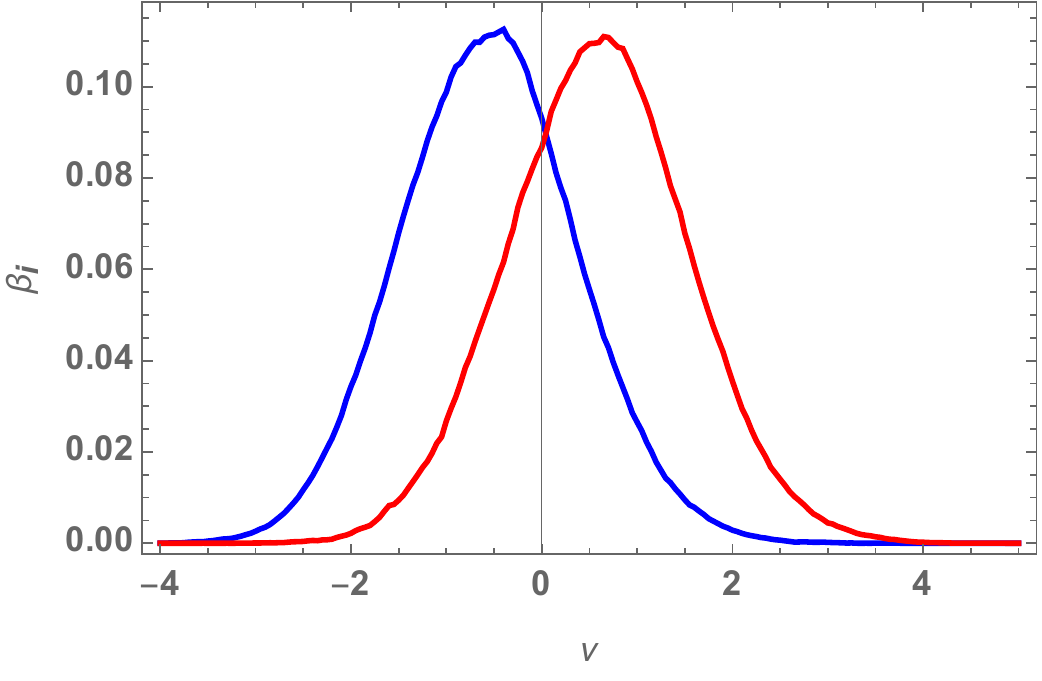}}
\end{subfigure}
\begin{subfigure}[b]{0.32\textwidth}
\caption{$f_{NL}=0.05$}
\mbox{\hskip 0.2cm\includegraphics[width=\textwidth]{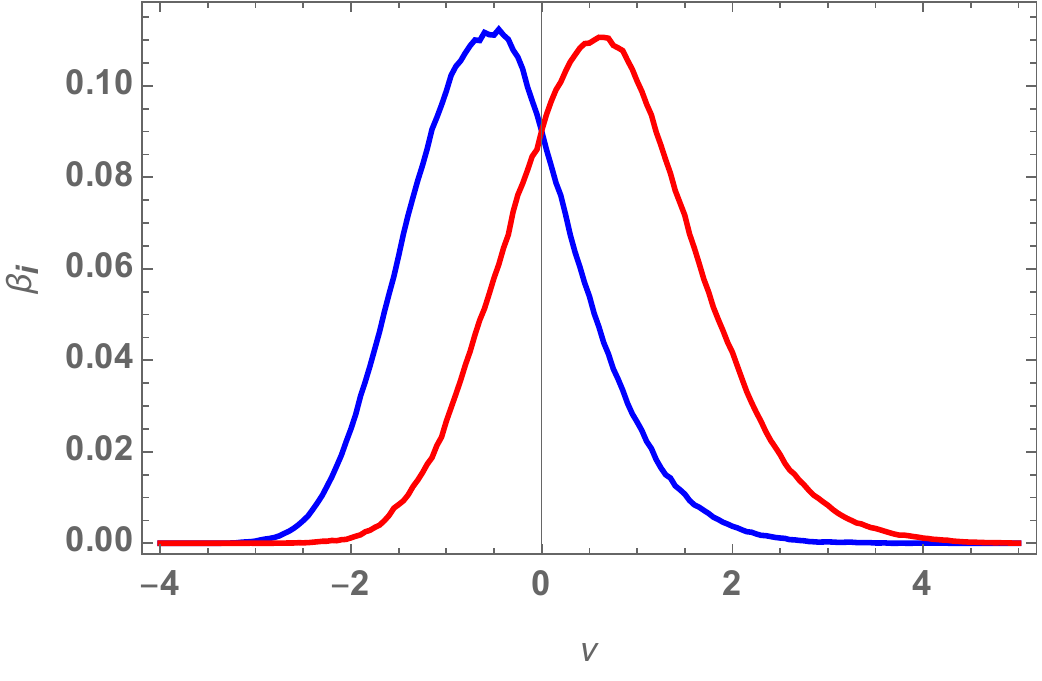}}
\end{subfigure}
\begin{subfigure}[b]{0.32\textwidth}
\caption{$f_{NL}=0.1$}
\mbox{\hskip 0.2cm\includegraphics[width=\textwidth]{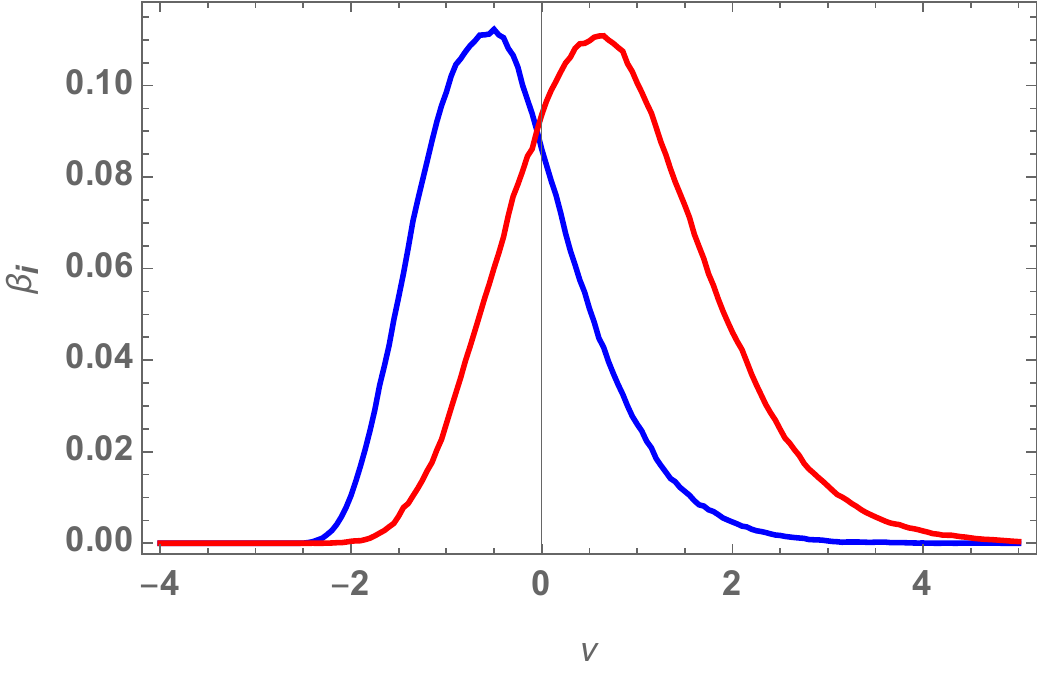}}
\end{subfigure}\\
\vspace{0.5cm}
\begin{subfigure}[b]{0.48\textwidth}
\mbox{\hskip -0.2cm\includegraphics[width=\textwidth]{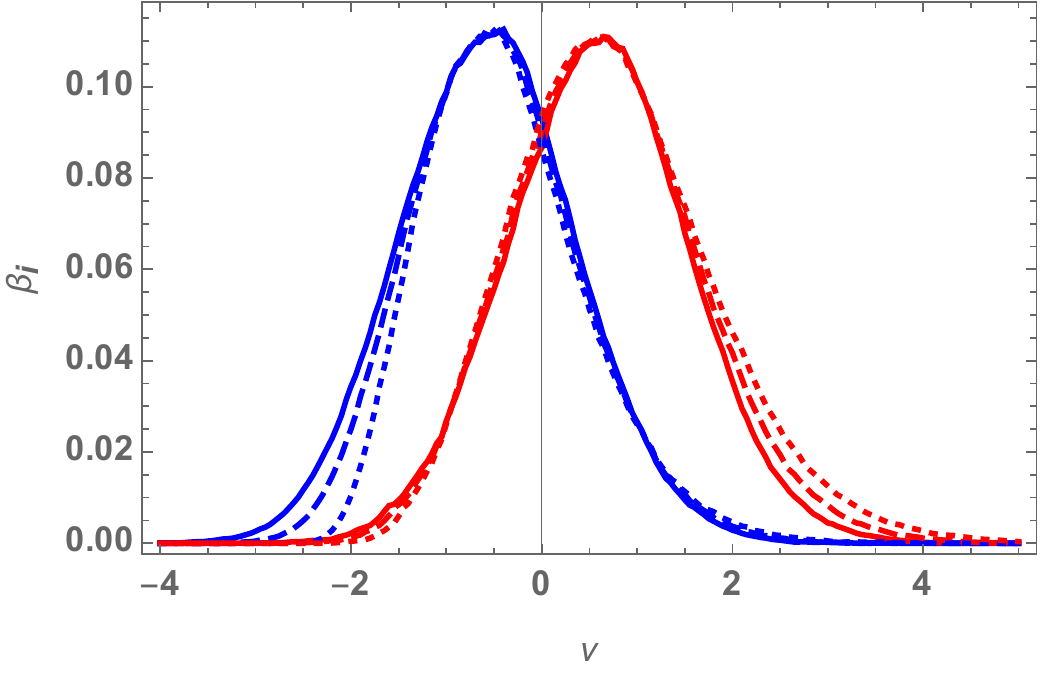}}
\end{subfigure}
\begin{subfigure}[b]{0.51\textwidth}
\mbox{\hskip 0.0cm\includegraphics[width=0.97\textwidth]{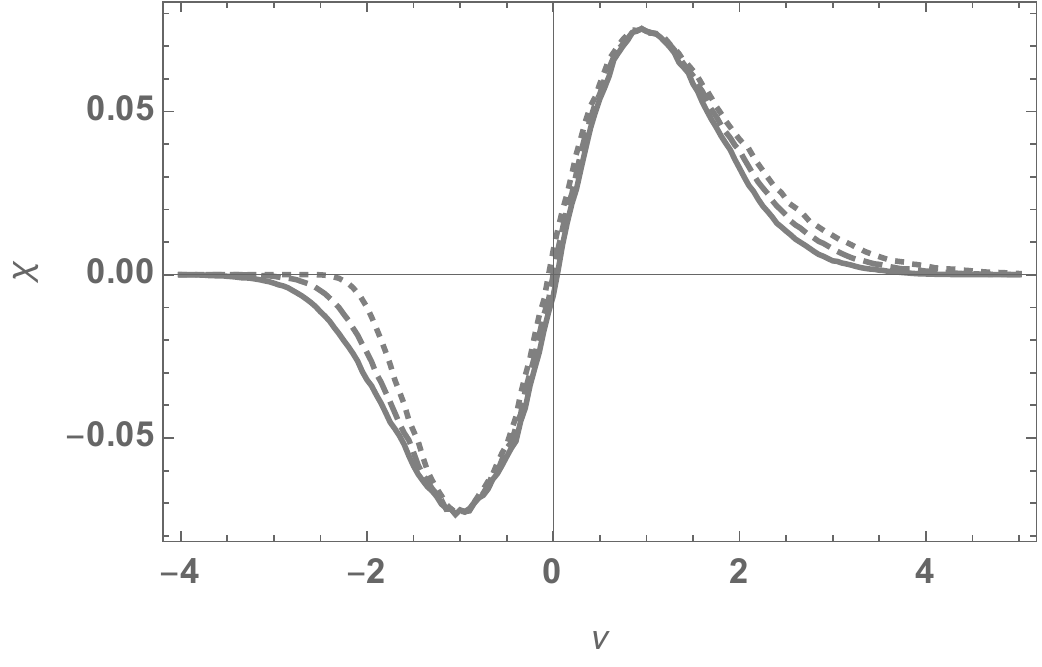}} 
\end{subfigure}
\caption{Betti Numbers \& Euler characteristic of non-Gaussian fields. For the same set of Gaussian and non-Gaussian random field realizations as in fig. \ref{fig:persBBKS} above, the panels show the Betti number curves $\beta_0(\nu)$ (red) and $\beta_1(\nu)$ (blue) and the corresponding Euler characteristic $\chi(\nu)$ (black), as a function of field threshold level $\nu=f/\sigma$.
  The non-Gaussian fields are of a local nature, characterized by the parameter $f_{NL}$. Top row: the Betti numbers $\beta_0$ (red) and $\beta_1$ (blue) for the individual random
  field realizations.
  Top left: Gaussian random field, with LCDM power spectrum (see text), and  $f_{NL}=0$ (by definition). Top centre: a random field with
  local non-Gaussianity parameter $f_{NL}=0.05$. Top right: a random field with local non-Gaussianity parameter $f_{NL}=0.1$. Note the characteristics shift of the Betti curves, manifesting itself as a systematic increase of skewness and (reduced) kurtosis (see table 3).
  Bottom row left: comparison of Betti curves $\beta_0$ (red) and $\beta_1$ (blue) of the 3 Gaussian and non-Gaussian random field realizations.
  Bottom row right: comparison between their Euler characteristic curves $\chi(\nu)$.
  Solid: Gaussian random field (LCDM power spectrum); dashed: local non-Gaussian field with $f_{NL}=0.05$; dotted: local non-Gaussian field with $f_{NL}=0.1$.}
\label{fig:bettinongauass}
\end{figure}

\bigskip
\noindent Within the context of primordial non-Gaussianities it is common to distinguish four different templates: local, enfolded, equilateral, and
orthogonal. These cover a large range of physical models of the early Universe. For an elaborate review of these templates see \cite{Yadav:2010, 2007JCAP...01..002C, 2014PDU.....5...75M, 2007JCAP...06..023C, 2005JCAP...09..011S}. In this study, we restrict our numerical study to the local template. A realization of a random field with local non-Gaussianities $f_{NG}:\mathbb{R}^2\to \mathbb{R}$ can be generated from a realization of a Gaussian random field $f_G:\mathbb{R}^2\to \mathbb{R}$ with zero mean $\mu=0$ and unit variance $\sigma^2=1$, using the non-linear transformation
\begin{align}
f_{NG}=f_G+f_{NL}f_G^2\text{,}
  \label{eqn:localNG}
\end{align}
with the non-Gaussianity parameter $f_{NL} \in \mathbb{R}$ \citep[see][]{Komatsu:2011}. In order to compare the Betti numbers and persistence diagrams for different non-Gaussianity parameters $f_{NL}$, we normalize $f_{NG}$ as follows, 
\begin{align}
f_{NG} \mapsto \frac{f_{NG} - \langle f_{NG}\rangle}{\langle f_{NG}^2 \rangle^{1/2}} \,. 
\end{align}
This  ensures that $f_{NG}$ is centered around zero, \textit{i.e.}, $\langle f_{NG}\rangle =0$,  and has unit variance,
\textit{i.e.}, $\langle f_{NG}^2\rangle =1$. For an illustration of a representative realization of a non-Gaussian field with local
non-Gaussianities with non-Gaussian parameter $f_{NL}=1$ see figure \ref{fig:NonGaus}. Note the asymmetry between the peaks and dips
in the field.

\begin{table}
\begin{center}
\begin{tabular}{ |p{2cm}||c|c|c|c|c|c|c|c|c|c| }
    \hline
                & \multicolumn{2}{c|}{}     & \multicolumn{2}{c|}{}                 & \multicolumn{2}{c|}{}            & \multicolumn{2}{c|}{}            & \multicolumn{2}{c|}{} \\
    $f_{NL}$    & \multicolumn{2}{c|}{$\mu$}     & \multicolumn{2}{c|}{$\sigma$}             & \multicolumn{2}{c|}{$\gamma_1$}     & \multicolumn{2}{c|}{$\gamma_2'$} & \multicolumn{2}{c|}{$\gamma_2 = \gamma_2'-3$} \\ 
                & \multicolumn{2}{c|}{mean}     & \multicolumn{2}{c|}{st. dev.}    & \multicolumn{2}{c|}{skewness}      & \multicolumn{2}{c|}{kurtosis}      & \multicolumn{2}{c|}{exces kurtosis}  \\ 
                 & \multicolumn{2}{c|}{}     & \multicolumn{2}{c|}{}                 & \multicolumn{2}{c|}{}            & \multicolumn{2}{c|}{}            & \multicolumn{2}{c|}{} \\
    \hline
    & $\beta_0$ & $\beta_1$ & $\beta_0$ & $\beta_1$ & $\beta_0$ & $\beta_1$ & $\beta_0$ & $\beta_1$ & $\beta_0$ & $\beta_1$   \\
    \hline
    & & & & & & & & & &  \\
    $0$        &    $0.64$ & $-0.55$     &    $0.95$ & $0.92$    &    $0.041$ & $0.085$ &  $3.03$ & $3.01$    &    $0.027$ & $0.015$    \\ 
    & & & & & & & & & &  \\
    $0.05$    &        $0.67$ & $-0.48$     &    $0.99$ & $0.88$  &     $0.25$ & $0.32$    &  $3.12$ & $3.18$      &    $0.12$ & $0.18$    \\ 
    & & & & & & & & & &  \\
    $0.1$        &    $0.76$ & $-0.38$     &    $1.05$ & $0.85$  &     $0.46$ & $0.64$    &  $3.25$ & $3.62$       &     $0.25$ & $0.62$    \\ 
    & & & & & & & & & &  \\
    \hline
  \end{tabular}
  \caption{Betti curve moments:  the mean, standard deviation, skewness and kurtosis of the curves $\beta_0(\nu)$ (the first entry) and $\beta_1(\nu)$ (the second entry) of a (LCDM) Gaussian random field, and two corresponding (local) non-Gaussian fields. The level of non-Gaussianity is specified by the parameter $f_{NL}$ (see expression~\eqref{eqn:localNG}).}
\label{table:nonGauss}
\end{center}
\end{table}

\medskip
In figure \ref{fig:persBBKS} we plot the persistence diagrams $\Pi_0$ and $\Pi_1$ as a function of parameter $f_{NL}$. The original random field $f_G$ was constructed with the cold dark matter power spectrum described by \cite{Bardeen:1986}. In the Gaussian case, \textit{i.e.}, $f_{NL}= 0$, the zero- and one-dimensional persistent diagrams are statistically symmetric, \textit{i.e.}, $\Pi_0(a,b) = \Pi_1(b,a)$. As $f_{NL}$ is increased, we see that this symmetry is broken.
The zero-dimensional persistent diagram $\Pi_0$ consists of many points in the upper right quadrant which are absent in the lower left quadrant of
one-dimensional persistent diagram $\Pi_1$. The presence of a small local non-Gaussian component leads to an observable effect in the persistence diagram
of the field. However, note that the scatter plots in particular emphasize the change in the outliers.

A more quantitative effect can be witnessed in the persistent Betti numbers (see figure \ref{fig:bettinongauass}) corresponding to the persistent
diagrams presented in figure \ref{fig:persBBKS}. For a Gaussian random field, \textit{i.e.}, $f_{NL}=0$, the Betti curves $\beta_0(\nu)$ and
$\beta_1(\nu)$ are bell-shaped curves. Also, the frames show that statistically the two Betti numbers are reflection symmetric, \textit{i.e.},
$\beta_0(\nu) = \beta_1(-\nu)$. In the non-Gaussian case, both symmetries are broken. The individual Betti curves $\beta_0(\nu)$, $\beta_1(\nu)$ get skewed
and the symmetry between the two curves is broken. Quantitatively, this effect can be seen in the behaviour of the mean, standard deviation, skewness and
kurtosis,

\begin{alignat*}{3}
\mu          &= E[\nu]\,;                 &&\sigma^2\,=\,E[(\nu - \mu)^2]\,;&& \nonumber\\
\gamma_1     &= E\left[\left(\frac{\nu - \mu}{\sigma}\right)^3\right]\,;\quad\quad&&
\gamma_2'     = E\left[\left(\frac{\nu - \mu}{\sigma}\right)^4\right]\,;\quad\quad&&\gamma_2\,=\,\gamma_2'-3\,.\nonumber
\end{alignat*}
The inferred values of these statistical moments are presented in table \ref{table:integrationvariables}. We observe that the mean $\mu$ of
both $\beta_0$ and $\beta_1$ increase for larger $f_{NL}$. Interestingly, we find a decrease of the standard deviation $\sigma$ of the Betti curve
$\beta_0(\nu)$ for increasing $f_{NL}$, while the standard deviation for the Betti curve $\beta_1(\nu)$ decreases. The skewness $\gamma_1$ and excess kurtosis
$\gamma_2$ also evolve non-trivially as a function of the non-Gaussianity parameter $f_{NL}$. A similar effect is present in
the Euler characteristic $\chi(\nu)$ (see figure~\ref{fig:bettinongauass}). Note that the Betti curves $\beta_i(\nu)$ completely determine
the Euler characteristic $\chi(\nu)$, \textit{i.e.},
\begin{align}
  \chi(\nu)\,=\,\beta_0(\nu)-\beta_1(\nu) + \beta_2(\nu)\,.
  \end{align}
Note that the second Betti number, $\beta_2$, vanishes. The Euler characteristic, and more generally the Minkowski functionals, of the CMB has been extensively studied in \cite{Eriksen:2004, Park:2004, Modest:2013, Buchert:2017, Ganesan:2017, Novaes:2016, Zhao:2014, Ducout:2013, Munshi:2012}. We believe that an analysis in terms of the two Betti curves and the persistence diagrams will have more statistical power to detect non-Gaussian deviations. For an initial investigation see \cite{Pranav:2019b}.

In summary, we see that the persistent homology of a random field is highly sensitive to the presence of non-Gaussianties. By studying the Betti curves of the cosmic microwave background radiation field, we can attempt to measure primordial non-Gaussianities. For a more detailed study of the persistent homology of non-Gaussianities on the two-dimensional sphere $\mathbb{S}^2$, we refer to the upcoming paper \cite{Feldbrugge:2020b}.

\section{Conclusion and Discussion}
\label{sec:conclusion}
The topology of superlevel set filtrations of Gaussian random fields has over the last decades received a lot of attention in terms of the study of the
Euler characteristic and the Minkowski functionals \cite{Eriksen:2004, Park:2004, Modest:2013, Buchert:2017, Ganesan:2017, Novaes:2016, Zhao:2014, Ducout:2013, Munshi:2012}. A major virtue of these functionals is that in the case of Gaussian random fields there
are analytical expressions for their expectation value \citep[see e.g][]{Pranav:2019a}. Deviations from the Gaussian nature of the primordial
temperature perturbations in the Cosmic Microwave Background are then expected to reveal themselves as differences with the Gaussian
expectation values. 

Potential deviations in the cosmic microwave background radiation field, resulting from primordial non-Gaussianities, can provide us with valuable clues about the physics of the early Universe \citep{guthpi1982,Eriksen:2004,Baumann:2009,Chen:2010,Komatsu:2011,Bartolo:2012}. However, the Euler characteristic and Minkowski functionals are summarizing
characteristics of the topology. Considerably
more topological information is contained in the homology of manifolds, in particular in their persistent homology. Homology groups for superlevel set filtrations
contain the complete information of the creation, merging and disappearance of topological features.

The present study focusses on the theoretical and analytical aspects of the homology of Gaussian random fields. The insights, understanding and
analytical expressions derived have the objective of providing a theoretical foundation for the numerical and data analysis studies within the
context of our project for developing and applying advanced concepts of topological data analysis \citep[see e.g.][]{Edelsbrunner:2009, Wasserman:2018} to the
description and quantification of the topology and of the large scale Universe. These range from the topology of the primordial
mass distribution up to the connectivity aspects of the complex spatial pattern of the cosmic web on Megaparsec scales \citep{Weygaert:2011, Park:2013,
  Pranav:2017,Pranav:2019a,Pranav:2019b,Nevenzeel:2019}. 

In this paper, we demonstrate that Betti numbers and persistence diagrams are a powerful tool for the more complete characterisation of the topology of the
primordial mass distribution \cite{Weygaert:2011,bachelorThesis,Pranav:2019a}. The study involves an analytical investigation of the topology of (Gaussian)
random fields based on Morse theory. Our
argument is based on Morse theory's principal topological thesis, the fact that the topology of a manifold is fully specified by the spatial distribution
and connectivity of the discrete set of its singularities  (maxima, minima and saddle points). We formalize the connectivity structure of the
field in terms of the Morse graph, which condenses the connectivity structure of the Morse-Smale complex into a graph network. In other words, it
introduces the concept of a graph to represent the connectivity of a manifold as specified by the Morse-Smale complex, the segmentation of space defined by the spatial distribution of the singularities and which consists of regions connecting minima and maxima via
the field's integral lines. Having compressed the topological information into the Morse graph, we proceed by a filtration process in terms
of the {\it incremental algorithm} for adding or removing topological features as new singularities in the Morse graph get included. In combination
with the analytically known field distribution functions for the maxima, minima, and saddles in a Gaussian random field, we are led to the
integral expressions for the Betti numbers and persistence diagrams of Gaussian random fields. By investigating the asymptotic behaviour of
the Betti numbers we are able to infer an accurate fitting function. Note that these analytical expressions have been
derived on the basis of a probabilistic calculation, and do not involve and resort to realizations of Gaussian random fields. 

Deviations from the obtained fitting formulas most likely indicate the presence of non-Gaussianities. We numerically demonstrate that both
the Betti numbers and the persistence diagrams are sensitive probes for the search for primordial non-Gaussianities in the CMB
\citep[also see][]{bachelorThesis,Chingangbam:2012,ColeShiu:2018,Pranav:2019b}. We evaluate both the Betti numbers and the persistent diagrams
for a range of non-Gaussian fields of the local template and demonstrate the existence of a systematic deformation of the distribution. We expect different non-Gaussianity templates to affect the distribution of the Betti numbers and
persistence diagrams in distinct ways. For a systematic analysis of the persistent homology of non-Gaussian fields, we refer to the upcoming
paper \cite{Feldbrugge:2020b}.

\section*{Acknowledgment}
We are particularly indebted to Robert Adler for his encouragement and illuminating discussions. We wish to thank Leon Doddema, whose help was
essential for solving key computer and data issues.  We are also grateful to Bernard Jones, Elisabetta Pallante, and Mathijs Wintraecken for useful
and insightful discussions, in particular at the early stages of this project. Job Feldbrugge thanks Neil Turok for enabling the continuation of research
on the presented study. Research at Perimeter Institute is supported by the Government of Canada through Industry Canada and by the Province of Ontario through
the Ministry of Economic Development, Job Creation and Trade. Rien van de Weygaert acknowledges support by the John Templeton Foundation grant nr. FP5136-O. He
is also very grateful for the hospitality of the Perimeter Institute, Waterloo, Canada, which allowed us to finish the present study. Pratyush Pranav acknowledges the support of the ERC advanced grand ARTHUS (no:740021).

\bibliographystyle{JHEP}

\bibliography{Homology}


\appendix

\section{Concepts of Homology theory}
\label{appendix:Homology}
This appendix presents a formal introduction to homology theory and Betti numbers. A more extensive treatment and discussion can be found in several books and
reviews \citep{Vegter:1997,Edelsbrunner:2009, Rote:2006, Robins:2015}. Appendices \ref{Simplices}, \ref{Chains}, \ref{PersistenceDiagrams} present the formal
definitions of simplicial complexes, chains and cycles and persistence diagrams. Appendix~\ref{Example} proceeds with a case study of a simple example of how to compute 
Betti numbers from first principle.

\subsection{Simplices}\label{Simplices}
\label{Simplical Complexes }
Networks and graphs are well-known concepts in mathematics. They describe the connections and relations between points or objects. 
In geometry they appear as tessellations, tilings of space composed of points, lines, surfaces and cells that in combination represent
a geometric object \citep[see e.g.][]{Okabe:2000}. They are known as a {\it Simplicial Complex}. A representative example of a simplicial complex
in two-dimensional
Euclidian space is shown in figure~\ref{fig:tes}. The points, lines, and surfaces are known as vertices, edges, and faces or,
in a more formal fashion, as 0-, 1- and 2-simplices. We write a simplex by listing its corresponding vertices (points), \textit{i.e.}, the line segment between point $A$ and $B$ is represented by the edge $AB$. The vertices $A$, $B$, and $C$ can span a face $ABC$. In the present study we only consider simplicial complexes embedded
in $2$-dimensional Euclidean space. However, simplicial complexes can be defined in any $D$-dimensional space.

Remark that the homology of a manifold can be captured by these simplices using a polyhedral decomposition of the manifold. The homology of a manifold is well-defined as it is independent of the chosen decomposition. In the subsequent sections we always assume a single simplicial decomposition of a manifold.

\subsection{Chains and the Boundary Operator}
\label{Chains}
A simplicial complex is composed of simplices: vertices, edges, and faces. Using a simplicial decomposition of the manifold $M$, we can construct the corresponding simplicial complex. Topology is interested in the global properties of the complex. We here we define a composite object known as a {\it chain}. On these chains we define the {\it boundary operator} which maps chains to their boundaries.\\

\subsubsection{Chain}
\noindent A $p$-chain is a collection of $p$-simplices of the complex. Formally a $p$-chain is a finite sum \\
\begin{align*}
c=\sum_\sigma c_\sigma \sigma,
\end{align*}
where $\sigma$ are $p$-simplices and $c_\sigma\in\mathbb{Z}/2\mathbb{Z}$\footnote{The group $\mathbb{Z}/2\mathbb{Z}$ can be seen as $\{0,1\}$ with the property that $1 + 0 = 1$ and $1+1=0$.}. The simplex $\sigma$ is included in the chain $c$ if and only if $c_\sigma=1$. The collection of all $p$-chains, corresponding to a simplicial decomposition of the manifold $M$, forms the chain group $C_p(M)$.

\subsubsection{Boundary \& Boundary Operator}
The ``boundary" of a $p$-simplex consists of a number of $(p-1)$-simplices. Given a $p$-simplex 
\begin{align}
\sigma = v_{i_0}v_{i_1}\cdots v_{i_p}
\end{align}
with vertices $v_i$, we define the boundary of the simplex as
\begin{align*}
&\partial_p(\sigma)=\sum^p_{j=0}v_{i_0}\cdots v_{i_{j-1}}v_{i_{j+1}}\cdots v_{i_p}\,,
\end{align*}
where the vertex $v_{i_j}$ has been dropped from the simplex. The zero-dimensional boundary operator $\partial_0$ is defined to be the zero-function, \textit{i.e.}, $\partial_0(v_i)=0$ for all vertices $v_i$. The boundary operator can be extended to chains assuming linearity, \textit{i.e.},
\begin{align}
\partial_p\left(\sum_\sigma c_\sigma \sigma\right) = \sum_\sigma c_\sigma \partial_p(\sigma)\,,
\end{align} 
making the boundary operator $\partial_p$ a map from the $p$-dimensional chain group $C_p$ to the $(p-1)$-dimensional chain group $C_{p-1}$.

\begin{figure}
\centering
\mbox{\hskip 1.5truecm\includegraphics[width=1.2\textwidth]{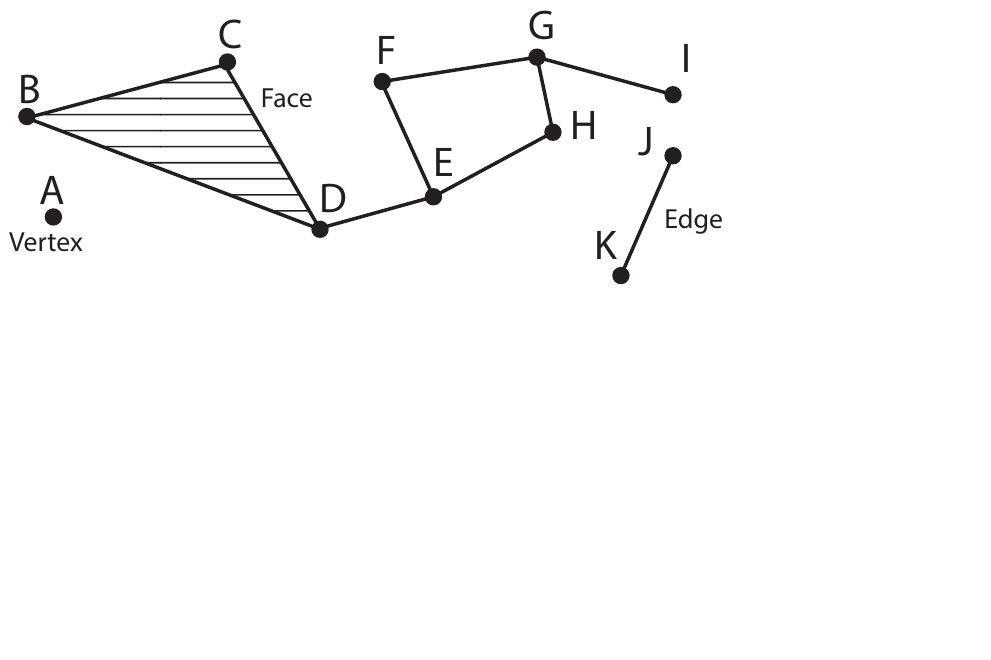}}
\vskip -7.0truecm
\caption{Simplicial Complex. An example of a simplicial complex, or tessellation, in two-dimensional Euclidean space. The letters mark the points known as vertices. The lines represent the edges and the shaded region represents a face.}
\label{fig:tes}
\end{figure}

\bigskip
\noindent The boundary operator takes a chain and returns the boundary of that chain. For illustrative purposes consider examples extracted from the simplicial complex in figure~\ref{fig:tes}. In these examples, we apply the
boundary operator on several chains in the simplicial complex. The boundary of the edge $JK$ consists of the boundary vertices $J$ and $K$, \textit{i.e.},
boundaries, \textit{i.e.}
\begin{align}
\partial_1(JK) = J + K\,.
\end{align}
The boundary operator returns an empty chain when operating on a loop or more formally a \textit{cycle}. Consider for example the boundary of the cycle $EF + FG + GH + EH$, \textit{i.e.},
\begin{align}
\partial_1(EF+FG+GH+EH) &= E + F + F + G + G + H + E + H\\
&=0\,.
\end{align}
By considering the face $BCD$ we observe that the boundary of a boundary vanishes
\textit{i.e.},
\begin{align}
\partial_1(\partial_2 (BCD)) &= \partial_1(CD + BD + BC)\\
&=  C + D + B + D + B + C\\
&=0\,.
\end{align}
The fact that the square of the boundary operator vanishes, \textit{i.e.},
\begin{equation}
  (\partial_{p-1}\circ{}\partial_{p})(c)=0\,.
\end{equation}
for a general $p$-chain $c$, can be proven from the definition of the boundary operator. In the next section we use this exactness property of chains to define the homology group.
  
\subsection{Homology Groups and Betti Numbers}
\label{PersistenceDiagrams}
Using the boundary operator acting on the chain groups corresponding to a simplicial decomposition of the manifold $M$, we construct an exact sequence
\begin{align*}
\dots\xrightarrow{\partial_{p+2}} C_{p+1}(M)\xrightarrow{\partial_{p+1}}C_p{M}\xrightarrow{\partial_{p}}C_{p-1}(M)\xrightarrow{\partial_{p-1}}\dots
\end{align*}
The sequence is exact since $\partial_{p-1}\circ{}\partial_{p}=0$. Using the boundary operator we construct two subsets of the chain group $C_p(M)$. We define the $p$-cycles $Z_p(M)$ and the $p$-boundaries $B_p(M)$ as the image and the kernel of the boundary operator, \textit{i.e.},
\begin{align*}
&Z_p(M)=\text{Ker }\partial_p\,,\\
& B_p(M)=\text{Im }\partial_{p+1}\,.
\end{align*}
The exactness ensures that $B_p(M)\subseteq C_p(M)$ is a linear subgroup of $Z_p(M)\subseteq C_p(M)$.

\bigskip
\noindent The $p^{\text{th}}$ homology group of $M$ is defined as the quotient group
\begin{align*}
H_p(M)=Z_p(M)/B_p(M).
\end{align*}
The group $H_p(M)$ should be seen as a partition of $Z_p(M)$ in sets (equivalence classes) that differ by an element in $B_p(M)$.

\bigskip
\noindent The $p^{\text{th}}$ Betti number of $M$ is defined as
\begin{align*}
\beta_p=\text{rank}\ H_p(M),
\end{align*}
which is the number of equivalence classes in $H_p(M)$. Note that for a compact manifold, the Betti numbers are finite.

\begin{figure}
\centering
\includegraphics[width=0.7\textwidth]{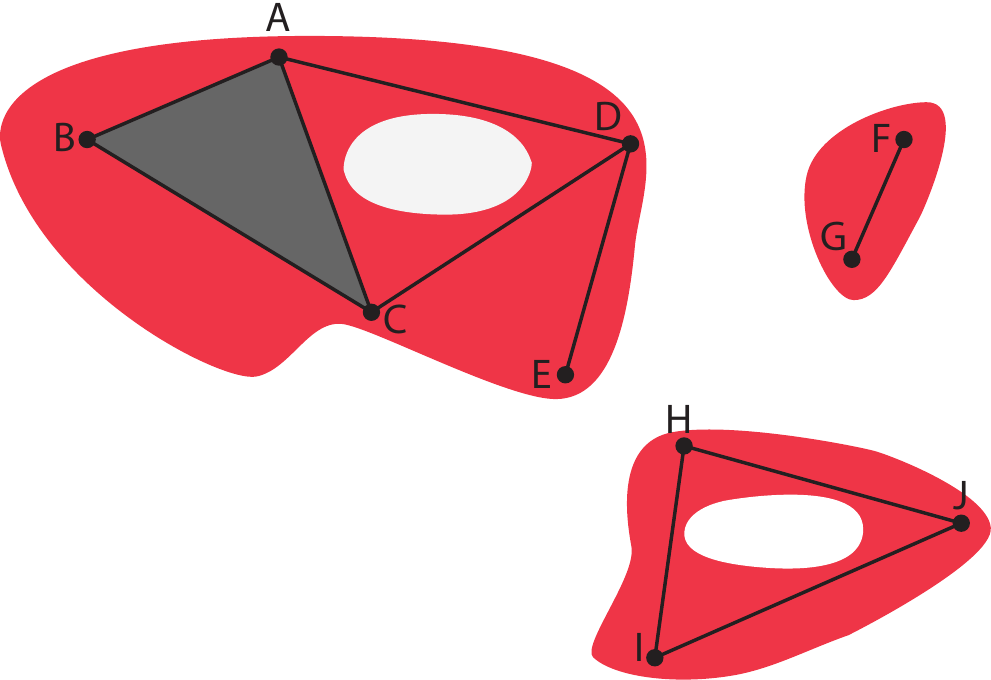}
\vspace{0.25cm}
\caption{Chain groups simplicial complex. A manifold is shown in red, with a specified collection
  of chains from the chain group.}
\label{fig:manifoldex}
\end{figure}
\subsection{Example}\label{Example}
On the basis of figure~\ref{fig:manifoldex}, we provide a visual illustration of the significance of the Betti numbers of
a manifold $M$ by following the method outlined above. 
Instead of working with a simplicial decomposition of the manifold, we consider the homology of the homologically equivalent simplicial representation denoted by the vertices, edges, and faces. 

The chain groups of the simplicial complex are given by the formal sums of the vertices, edges and faces present in the simplicial complex:
\begin{align*}
\mathcal{C}_0&=\{A,B,C,D,E,F,G,H,I,J\}\\
\mathcal{C}_1&=\{AB,AC,AD,BC,CD,DE,FG,HI,HJ,IJ\}\\
\mathcal{C}_2&=\{ABC\},
\end{align*}
where we denote a $p$-simplex by its corresponding vertices.
For these chain groups the boundary groups are generated by the chains:
\begin{align*}
\mathcal{B}_0=&\{A+B,A+C,A+D,B+C,C+D,D+E,F+G,H+I,H+J,I+J\}\\
\mathcal{B}_1=&\{AB,AC,BC\}\\
\mathcal{B}_2=&\{0\},
\end{align*}
since $\partial_1 (AB)=A+B$, $\partial_1(AC)=A+C$ etc. The cycle groups are generated by the chains:
\begin{align*}
\mathcal{Z}_0&=\mathcal{C}_0\\
\mathcal{Z}_1&=\{AB+AC+BC,AC+AD+CD,HI+HJ+IJ\}\\
\mathcal{Z}_2&=\{0\},
\end{align*}
since $\partial_1(AB+AC+BC)=A+B+A+C+B+C=0$ etc. We evaluate homology groups by partitioning $\mathcal{Z}_i$ into subsets whose elements differ by an element generated by $\mathcal{B}_i$ for $i=0,1,2$. The homology groups are given by
\begin{align*}
&H_0=\{A,F,H\}\\
&H_1=\{AC+AD+CD,HI+HJ+IJ\}\\
&H_2=\{0\}\,,
\end{align*}
where we represent the equivalence classes by a single element of the class. As the Betti numbers are the number of equivalence classes in the homology groups, we find the Betti numbers
\begin{align*}
&\beta_0=3,\ \beta_1=2\text{ and }\beta_2=0\,.
\end{align*}
This is what one would expect, as the manifold $M$ has three connected components and two holes. 
\end{document}